\documentclass[sigconf, nonacm]{acmart}

\usepackage{amsmath,epsfig}
\usepackage{cleveref}

\crefformat{section}{\S#2#1#3} 
\crefformat{subsection}{\S#2#1#3}
\long\def\comment#1{}

\usepackage{makecell}
\usepackage{color}
\usepackage{tikz}
\usepackage{bm}
\usepackage{graphicx}
\usepackage{xspace}
\usepackage{subfigure}
\usepackage{caption}
\usepackage{amsmath}
\usepackage{float}
\usepackage{graphicx}
\usepackage{url}
\usepackage{epsfig}
\usepackage{mathrsfs}
\usepackage{multirow}
\usepackage{color}
\usepackage{epstopdf}
\usepackage{epsf}
\usepackage[ruled,vlined, noend]{algorithm2e}
\usepackage{listings}
\usepackage[nounderscore]{syntax}
\usepackage{setspace}
\usepackage{framed}
\usepackage{nopageno}
\usepackage{hyperref}
\usepackage{mathtools}
\usepackage{physics}
\usepackage{pifont}
\usepackage{fancybox}
\usepackage{colortbl}
\usepackage{balance}

\newcounter{example}[section]
\renewcommand{\theexample}{\nthesection.\arabic{example}}
\newenvironment{example}{
     \refstepcounter{example}
     {\vspace{1ex} \noindent\bf  Example  \theexample:}}{
     \vspace{1ex}} 

\newcounter{definition}[section]
\renewcommand{\thedefinition}{\nthesection.\arabic{definition}}

\newcounter{theorem}[section]
\renewcommand{\thetheorem}{\nthesection.\arabic{theorem}}

\newcounter{lemma}[section]
\renewcommand{\thelemma}{\nthesection.\arabic{lemma}}

\newcounter{remark}[section]
\renewcommand{\theremark}{\nthesection.\arabic{remark}}

\newcommand{\nthesection}{\arabic{section}}

\newcommand{\stitle}[1]{\vspace{1ex} \noindent{\bf #1}}
\newcommand{\etitle}[1]{\vspace{1ex} \noindent{\underline{\em #1}}}

\newcommand{\kw}[1]{{\ensuremath {\mathsf{#1}}}\xspace}

\newcommand*{\Scale}[2][4]{\scalebox{#1}{$#2$}}

\newcommand{\beqn}{\begin{eqnarray*}}
\newcommand{\eeqn}{\end{eqnarray*}}

\newcounter{ccc}

\newcommand{\kwnospace}[1]{{\ensuremath {\mathsf{#1}}}}

\newcommand{\cmark}{\ding{51}}%
\newcommand{\xmark}{\ding{55}}%

\newcommand{\NTK}{{\sl Neural~Tangents}\xspace}
\newcommand{\JAX}{{\sl JAX}\xspace}

\newcommand{\GBDT}{{$\texttt{GBDT}$}\xspace}
\newcommand{\NN}{{$\texttt{NN}$}\xspace}

\newcommand{\TLSTM}{{$\texttt{TLSTM}$}\xspace}
\newcommand{\MSCN}{{$\texttt{MSCN}$}\xspace}
\newcommand{\NNGP}{{$\texttt{NNGP}$}\xspace}
\newcommand{\GPRBF}{{$\texttt{GP-RBF}$}\xspace}
\newcommand{\Postgres}{{$\texttt{Postgres}$}\xspace}
\newcommand{\XGBoost}{{\sl XGBoost}\xspace}
\newcommand{\DBEst}{{\sl DBEst}\xspace}
\newcommand{\DeepDB}{{\sl DeepDB}\xspace}
\newcommand{\Naru}{{\sl Naru}\xspace}
\newcommand{\NeuroCard}{{\sl NeuroCard}\xspace}
\newcommand{\DeepDBExp}{{$\texttt{DeepDB}$}\xspace}
\newcommand{\NeuroCardExp}{{$\texttt{NeuroCard}$}\xspace}
\newcommand{\Qerror}{\kwnospace{q}\mbox{-}\kw{error}}
\newcommand{\DeepEnsemble}{{\sl Deep~Ensemble}\xspace}

\newcommand{\DeepEns}{{$\texttt{DeepEns}$}\xspace}
\newcommand{\BNNMCD}{{$\texttt{BNN-MCD}$}\xspace}

\newcommand{\rdbm} {{\sl RDBMS}\xspace}

\newcommand{\IN}{\kw{IN}}

\newcommand{\forest}{\kw{forest}}

\newcommand{\higgs}{\kw{higgs}}
\newcommand{\tpch}{\kwnospace{TPC}\mbox{-}\kw{H}}
\newcommand{\tpcds}{\kwnospace{TPC}\mbox{-}\kw{DS}}

\newcommand{\Norm}{\mathcal{N}}
\newcommand{\Real}{\mathbb{R}}
\newcommand{\Zero}{\bm{0}}
\newcommand{\Ident}{\mathcal{I}}
\newcommand{\Cov}{\mathsf{K}}
\newcommand{\Expect}{\mathbb{E}}
\newcommand{\Variance}{\mathbb{V}}
\newcommand{\Dist}{\mathcal{D}}
\newcommand{\act}{\sigma}
\newcommand{\loss}{\mathcal{L}}
\newcommand{\relu}{\mathsf{ReLU}}
\newcommand{\errFunc}{\mathsf{Erf}}
\newcommand{\softmax}{\mathsf{softmax}}
\newcommand{\sigmoid}{\mathsf{sigmoid}}
\newcommand{\COUNT}{\kw{cnt}}
\newcommand{\AVG}{\kw{avg}}
\newcommand{\SUM}{\kw{sum}}

\begin{document}
\settopmatter{printacmref=false}
\fancyhead{}
\fancyfoot{}
\title{Uncertainty-aware Cardinality Estimation \\ by Neural Network Gaussian Process}
\pagestyle{empty}

\author{Kangfei Zhao}
\affiliation{%
  \institution{The Chinese University of Hong Kong}
}
\email{{kfzhao}@se.cuhk.edu.hk}

\author{Jeffrey Xu Yu}
\affiliation{%
  \institution{The Chinese University of Hong Kong}
}
\email{yu@se.cuhk.edu.hk}

\author{Zongyan He}
\affiliation{%
  \institution{Renmin University of China}
}
\email{hzy9819@ruc.edu.cn}

\author{Hao Zhang}
\affiliation{%
  \institution{The Chinese University of Hong Kong}
}
\email{hzhang@se.cuhk.edu.hk}

\def\thepage{\arabic{page}}


\begin{abstract}

Deep Learning (DL) has achieved great success in many real
applications. Despite its success, there are some main problems when
deploying advanced DL models in database systems, such as
hyper-parameters tuning, the risk of overfitting, and lack of prediction 
uncertainty. In this paper, we study cardinality
estimation for SQL queries with a focus on uncertainty, which we
believe is important in database systems when dealing with a large
number of user queries on various applications. With uncertainty
ensured, instead of trusting an estimator learned as it is, a query
optimizer can explore other options when the estimator learned has a 
large variance, and it also becomes possible to update the estimator
to improve its prediction in areas with high uncertainty.
The approach we explore is different from the direction of deploying
sophisticated DL models in database systems to build cardinality
estimators. We employ Bayesian deep learning (BDL), which serves as a
bridge between Bayesian inference and deep learning. The prediction
distribution by BDL provides principled uncertainty calibration for
the prediction. In addition, when the network width of a BDL model
goes to infinity, the model performs equivalent to Gaussian Process
(GP). This special class of BDL, known as Neural Network Gaussian
Process (NNGP), inherits the advantages of Bayesian approach while
keeping universal approximation of neural network, and can utilize a
much larger model space to model distribution-free data as a
nonparametric model. We show that our uncertainty-aware NNGP estimator
achieves high accuracy, can be built very fast, and is robust to query
workload shift, in our extensive performance studies by comparing with
the existing approaches.


\end{abstract}

\maketitle

\section{Introduction}

\begin{table*}[!t]
{ \footnotesize 
\caption{ML/DL Approaches for AQP and Cardinality Estimation}
\label{tbl:exist_approaches}
\vspace*{-0.3cm}
\begin{center}
    \begin{tabular}{|l l l c c l l @{} c|} \hline
    \multirow{2}{*}{\bf Approach}  & \multicolumn{4}{c}{\bf Supported
      Queries} & \multirow{2}{*}{\bf Learning Strategy} &
    \multirow{2}{*}{\bf Model} & \multirow{2}{*}{\bf Model Update}
    \\ \cline{2-5} 
    \multicolumn{1}{|l}{} & {\bf Join} & {\bf Selection} & {\bf
      Aggregate} & {\bf Group By} & \multicolumn{3}{l|}{}  
     \\\hline\hline
\DBEst~\cite{DBLP:conf/sigmod/MaT19}  & precomp. join & num., cate. &
\COUNT,\SUM,\AVG,etc. & \cmark & supervised \& unsupervised & KDE \& GBDT  & \xmark
\\ \hline
\DeepDB~\cite{DBLP:journals/pvldb/HilprechtSKMKB20} & precomp. join &
num., cate. & \COUNT,\SUM,\AVG & \cmark & unsupervised & SPN & \cmark \\ \hline
Thirumuruganathan et. al.~\cite{DBLP:conf/icde/Thirumuruganathan20}  & \xmark
& num., cate.& \COUNT,\SUM,\AVG & \cmark & unsupervised & VAE & \xmark \\\hline
\hline
Kiefer et al.~\cite{DBLP:journals/pvldb/KieferHBM17} & Equi-join &
num., cate. & \COUNT & \xmark & unsupervised & KDE & \cmark
\\ \hline
Kipf et al.~\cite{DBLP:conf/cidr/KipfKRLBK19} & PK/FK join & num.,
cate. & \COUNT & \xmark & supervised  & MSCN & \xmark \\\hline
Dutt et al.~\cite{DBLP:journals/pvldb/DuttWNKNC19} & \xmark &
num., cate. & \COUNT & \xmark & supervised & NN, GBDT & \xmark
\\ \hline
Dutt et al.~\cite{dutt2020efficiently} & PK/FK join &
num., cate. & \COUNT & \xmark & supervised & GBDT & \cmark
\\ \hline
Sun et al.~\cite{DBLP:journals/pvldb/SunL19} & PK/FK join & num.,
cate., str. & \COUNT & \xmark & supervised & Tree LSTM & \cmark
\\ \hline
Hasan et al.~\cite{DBLP:conf/sigmod/HasanTAK020} & \xmark & num.,
cate. & \COUNT & \xmark & unsupervised & MADE & \xmark \\ \hline
\Naru~\cite{DBLP:journals/pvldb/YangLKWDCAHKS19} & precomp. join &
num., cate. & \COUNT & \xmark & unsupervised & MADE, Transformer &
\cmark \\ \hline 
\NeuroCard~\cite{DBLP:journals/corr/abs-2006-08109} & Full outer join & num., cate., & \COUNT & \xmark & unsupervised & MADE, Transformer & \cmark \\ \hline
\end{tabular}
\end{center}
}
\vspace*{-0.2cm}
\end{table*}

The learning approaches are shifting from the traditional ML (Machine
Learning) models (e.g., KDE, GBDT) to DL (Deep Learning) models (e.g.,
NN, MSCN, VAE, MADE, Transformer, SPN, LSTM).  The shifting is
motivated by the powerful approximation capability of neural networks
in end-to-end applications and the high efficiency of DL
frameworks. Compared with traditional ML, DL has achieved a great
improvement on estimation accuracy, and more and more advanced DL
architectures are devised to improve the performance of the learning
tasks with deeper layers to pursue more powerful modeling capability.
In database systems, the DL models have been extensively studied for
query optimization~\cite{DBLP:journals/pvldb/MarcusNMZAKPT19}, index
recommendation~\cite{DBLP:conf/sigmod/DingDM0CN19}, view
materialization~\cite{DBLP:journals/corr/abs-1903-01363}, and
cardinality
estimation~\cite{DBLP:journals/pvldb/KieferHBM17,DBLP:conf/cidr/KipfKRLBK19,DBLP:journals/pvldb/DuttWNKNC19,dutt2020efficiently,DBLP:journals/pvldb/SunL19,DBLP:conf/sigmod/HasanTAK020,DBLP:journals/pvldb/YangLKWDCAHKS19,DBLP:journals/corr/abs-2006-08109}.
Despite the success of DL approaches, the complex DL approaches
incur some main problems in general.

The first is hyper-parameters, on which the performance of DL models
highly relies, including the training hyper-parameters and network architecture
configurations. Note that onerous effort of parameter tuning must be
paid to pursuing satisfying performance, which is labor-intensive.
Although autoML tools~\cite{hutter2019automated} have been developed
to avoid human-in-the-loop, deploying such tools in database systems
needs great effort as searching a suitable hyper-parameter combination
needs a huge number of trials with high computation cost. It is worth
mentioning that a well-tuned model for one database is difficult to
transfer to other databases, which means that the retraining/tuning
needs repeating when the underlying data changes significantly.

The second is a high risk of overfitting that DL models are exposed to.
Subject to a particular family of function designed, general DL models
are indexed by a large number of parameters, fully fitted by the
training data. It is assumed testing data is from the same underlying
distribution of the training data, otherwise the prediction may have a 
large variance.
Many regularization techniques can be used, but they can only
alleviate the problem to some degree.
Collecting more data
to enhance training helps to reduce the variance. But this means a 
higher cost for acquiring the data/ground truth and training.
%
%
%

The third is prediction belief, which is an important issue we focus
on in this work.
The issue is that DL models cannot capture and convey their prediction
belief, namely, how probably their prediction is accurate or how much
is the prediction uncertainty.  The uncertainty comes from different
sources, for example, the noise of the training data, the
dissimilarity of the test data from the training data, the mismatching
of the model class to the data to be described.  For classification
tasks, DL models predict the probability distribution that one data
point is associated with the candidate classes by a $\softmax$
function, which is shown to be over-confident on the most likely class
\cite{DBLP:conf/icml/GuoPSW17}.  For regression tasks (e.g.,
cardinality estimation), DL models can only output a scalar value
without any uncertainty measurements of the prediction, such as
variance and confidential interval.  It is highly desirable to avoid 
situations where we have no choices but trust the DL predictions being
made in database systems. In other words, as a database system to
support a large number of users in various applications, where user
queries may be different and databases will be updated from time to
time,
%
%
what database systems require is not only an accurate model
prediction, but also an indication of how much the predictions can be
trusted regarding the learned model.

In this paper, we study cardinality estimation for SQL queries with
selection, projection and join.  We focus on
uncertainty. The approach we take can also address the hyper-parameter
and risk of overfitting.
Table~\ref{tbl:exist_approaches} summarizes the recent ML/DL approaches
studied for AQP (Approximate Query Processing) (the top three) 
and cardinality estimation (the bottom eight).
These approaches are categorized into supervised learning approaches
and unsupervised learning approaches.  The supervised learning
approaches are query-driven, which learn a function that maps the
query features to its cardinality.  The unsupervised learning
approaches are data-driven, which learn the joint probability distribution of the
underlying relational data.  The supervised learning approaches are
easy to deploy with relatively low training and prediction overhead,
however, they lack robustness to shifting query workloads.  On 
the contrary, the unsupervised learning approaches are robust to
different query workloads.  But, building a model consumes many 
resources as large volumes of data need to be fed into the model
multiple times.  For example, for a single relation within 100MB, the
unsupervised learning estimators
\DeepDB~\cite{DBLP:journals/pvldb/HilprechtSKMKB20} and
\NeuroCard~\cite{DBLP:journals/corr/abs-2006-08109} take hundreds and
thousands of seconds to train a model on a 56-core CPU, which is
2-3 orders slower than the lightweight supervised learning
estimators~\cite{DBLP:journals/pvldb/DuttWNKNC19}.  Furthermore, the
unsupervised estimators suffer from underestimation for range queries
as the prediction is conducted by integration over the learned
distribution regarding the query condition.
In this work, we study cardinality estimation based on a supervised
approach with an uncertainty guarantee.
Consider cardinality estimation for an SQL query. If a DL-based
estimator provides an estimate without uncertainty, the optimizer used
can only take it as it is. If a DL-based estimator provides an
additional probability distribution with mean, say $\mu$, and standard
deviation, say $2\mu$, as its predictive cardinality. Then, the query
optimizer can take it as inaccurate since the coefficient of variation
is up to 2, and the optimizer can either explore other options or use
the built-in estimator for query planning.
It is important that the system is aware that the estimator may be
outdated and need retraining, if the learned estimator delivers a low
confidence level for a large fraction of arrival queries.
Till now, all the DL-based cardinality approaches cannot provide the
prediction uncertainty.

\begin{figure}[t]
\centering
\includegraphics[width = 0.8\columnwidth]{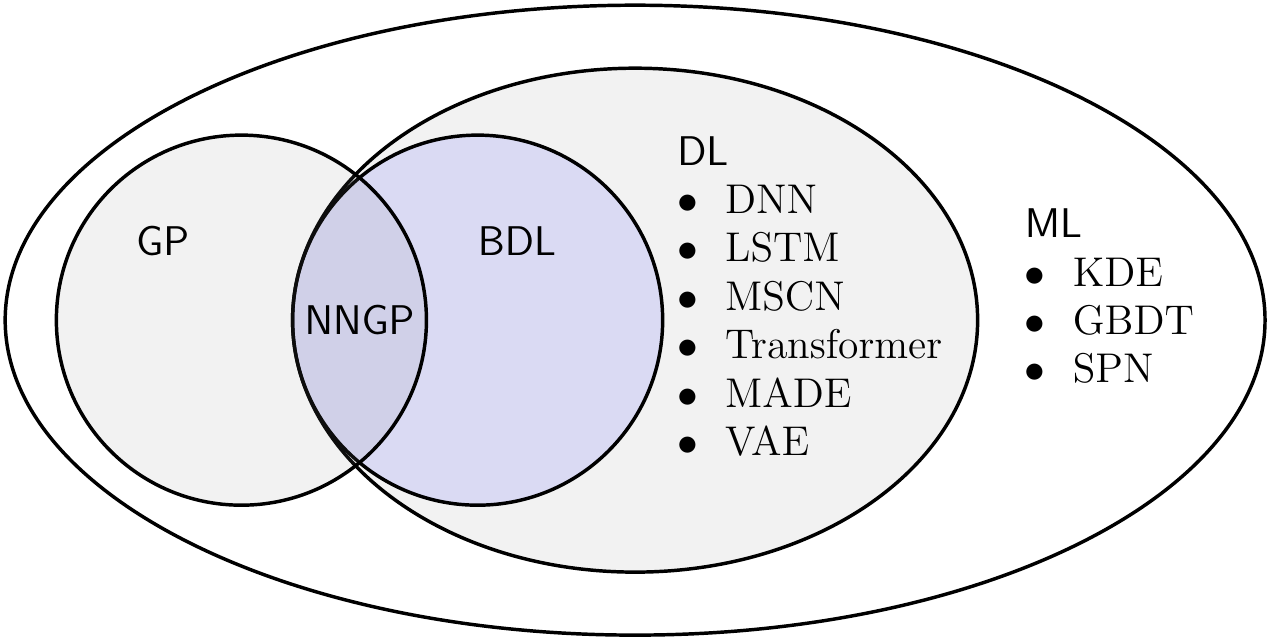}
\vspace*{-0.2cm}
\caption{From ML/DL estimators To NNGP (GP \& BDL)}
\label{fig:nngp:relationship}
\end{figure}

In this work, different from the direction of deploying sophisticated
DL models in database systems, we employ Bayesian deep learning (BDL)
to build cardinality estimators, where BDL serves as a bridge between
Bayesian inference and deep learning~\cite{DBLP:journals/tkde/WangY16}.
In a nutshell, BDL imposes a
prior distribution on the neural network weights, and the derived
models are ensembles of neural networks in a particular function space, which
output a distribution as the prediction.
The prediction distribution provides principled uncertainty
calibration for the prediction.  This uncertainty helps the model to
improve itself explicitly, by collecting the information of testing
data with higher uncertainty and retraining/fine-tuning the model.  In
addition, the prediction distribution by BDL is robust to the
overfitting problem.
The parameters and hyper-parameters of BDL models are much fewer
(i.e., the statistics of the prior distribution and the prior type,
respectively), in versus to the parameters and hyper-parameters that
need in DL models (i.e., the weights of the neural networks and the
network architecture configurations, number of layers, number of
hidden units in each layer), and in DL training configurations (i.e.,
optimization algorithm, learning rate, batch size, epochs, etc.).
When the network width of the BDL model goes to \emph{infinity},
the model performs equivalent to Gaussian Process
(GP)~\cite{neal1996priors}, named Neural Network Gaussian Process (NNGP).
Exact Bayesian inference can be used
to train this special GP as a lightweight cardinality estimator,
while offering a more powerful generalization capability than a finite
wide neural network.
NNGP keeps the flexible modeling capability of deep learning, while
offering robustness and interpretability by Bayesian inference.
Fig.~\ref{fig:nngp:relationship} delineates the extraordinary
standpoint of our NNGP estimator among the existing ML/DL-based
estimators in Table~\ref{tbl:exist_approaches}, for cardinality
estimation. Here, KDE and GBDT are classical ML models while SPN is a
new type of probabilistic graphical model with deeper layers. DL
models are associated with different neural network architectures.  A
BDL model is a special kind of DL model in which the neural network
parameters are probability distributions. We explore a new model,
named NNGP, which is the BDL model with infinite wide hidden layers,
and can be built using GP.

\stitle{Contribution.} The main contributions of this paper are
summarized as follows.  \ding{202} We employ an advanced BDL model,
named Neural Network Gaussian Process (NNGP), to build cardinality
estimators for relational database systems.  Our NNGP estimator can
support range queries on single relations and equi/theta-join on
multiple relations for a database.  To the best of our knowledge, this
is the first exploration of BDL approaches for database applications.
\ding{203} As the first DL-based cardinality estimator which supports
principled uncertainty estimation, we investigate the uncertainty that
NNGP captures, compared with existing uncertainty quantification
approaches for DL.  \ding{204} We conduct extensive experimental
studies to verify the effectiveness and efficiency of NNGP estimator via
comparison with recent ML/DL based estimators.  The NNGP estimator
 distinguishes from existing estimators by its swift model
construction and prediction, robustness to workload shifts, and 
appealing accuracy.

\stitle{Roadmap.} \cref{sec:ps} gives the problem statement.
In~\cref{sec:ii}, we give an overview of the NNGP
estimator, and introduce Bayesian learning,
GP and NNGP in \cref{sec:bg}.  Then, we elaborate on the uncertainty calibration of NNGP 
in \cref{sec:uncertainty}. ~\cref{sec:exp} reports the
experimental results.  Finally, we review the related works
in~\cref{sec:rw} and conclude the paper in~\cref{sec:conclusion}.

\section{Problem Statement}
\label{sec:ps}

A relational database consists of a set of relations, $\{ R_1, R_2,
\cdots, R_N\}$, where a relation $R_i$ has $d_i$ attributes such as
$R_i = (A_{1}, \cdots A_{d_i})$. Here, an attribute $A_j$ is either a
numerical attribute in a domain with given range $[min_j, max_j]$ or a
categorical attribute with a discrete finite domain $\{c_1, c_2,
\cdots c_{m_j}\}$.

We study cardinality estimation for the select-project-join SQL
queries with conjunctive conditions. 
A selection on an attribute is either a range filter (i.e., $[lb_j,
  ub_j]$, denoting the condition $lb_j \leq A_{j} \leq ub_j$), if the
attribute is numerical, or \IN filter (i.e., $A_j~\IN~C $, $C \subset
\{ c_1, c_2, \cdots c_{m_j}\}$, denoting the condition $\exists c_k \in C,
A_j = c_k$), if the attribute is categorical.
%
%
%
A projection can be on any attributes. A join can be
equi/theta-join. For joining over numerical attributes, the join
condition can be $\{<, \leq, =, >, \geq, \neq\}$. For joining over
categorical attributes, the join condition is either $=$ or $\neq$, as
the categorical domain is order-free. The primary-foreign key join, or
PK/FK-join, is treated as a special equi-join with the extra
constraints. We also support self-joins, which are conducted by
renaming the same relations. An example is given below.
\begin{align}
\sigma_{(100 \leq R_1.A_1 \leq 200) \land (R2.A_2~\IN~\{ 15, 20, 25
  \})} \displaystyle{ (R_1 \mathop{\Join}_{\substack{R_1.A_3
      \leq R_2.A_3}} R_2)} \nonumber   
\end{align}
where $\sigma$ is the select operator and $\Join$ is a join operator.
It is worth mentioning that we support general select-project-join SQL queries.
All the existing learned estimators do not support joins other than PK/FK joins.  
For example, \cite{DBLP:journals/corr/abs-2006-08109} does not support cyclic join queries and selection conditions on join attributes and \cite{DBLP:journals/pvldb/HilprechtSKMKB20} does not support multiple selection conditions on join attributes. 

The cardinality of an SQL query $q$ is the number of resulting tuples,
denoted as $c(q)$. To learn a cardinality estimation, like the
existing work, we require a set of joinable attribute pairs $\{
(R_1.A_i, R_2.A_j),$ $\cdots \}$, in addition to a set of relations
$\{ R_1,\cdots R_N\}$.  For example, two relations $\{R_1, R_2\}$ with
joinable attribute pairs $\{(R_1.A_3, R_2.A_3)\}$.

\stitle{The Problem Statement}: Given a set of relations and a set of
joinable attribute pairs, learn a model $\mathcal{M}: q \mapsto
\mathbb{R}$ from a training query set $\{ (q_1, c(q_1)),$ $(q_2,
c(q_2)), \cdots\}$, where $q_i$ is an SQL select-project-join query
and $c(q_i)$ is its actual cardinality, to predict the cardinality for
unseen queries.

We use \Qerror to evaluate the accuracy of the
estimated value.
\begin{equation}
\Qerror(q) = max\bigg\{\frac{c(q)}{\hat{c}(q)}, \frac{\hat{c}(q)}{{c}(q)} \bigg\}  
\end{equation}
Intuitively, \Qerror quantifies the factor by which the estimated
count ($\hat{c}(q_i)$) differs from the true count ($c(q_i)$).  It is
symmetrical and relative so that it provides the statistical stability
for true counts of various magnitudes. Here, we assume $c(q) \geq 1$
and $\hat{c}(q) \geq 1$.

We learn an uncertainty-aware model, and we do not make any assumptions
on data and query, e.g., the distribution of attribute values, the
independence of attributes and relations, the distribution of the
selection/join conditions of the queries, etc.

\stitle{Query Encoding}: Following the existing work, a
select-project-join query we support can be encoded by a fixed length
vector. The encoding consists of two parts: the selection conditions
and the join conditions. The two parts are encoded separately and
concatenated as follows.
\[
\Scale[1.2]{
\overbrace{\underbrace{~0.25~0.5~}_{R_1.A_1}~\cdots~}^{R_1~\text{selections}}\overbrace{\underbrace{~1~1~0~1~0~}_{R_2.A_2}~\cdots~}^{R_2~ \text{selections}}\overbrace{\underbrace{~1~1~0~}_{(R_1.A_3, R_2.A_3)}~0~0~0~
}^{\text{ joins}}
}
\]
For selection condition, the encoding of all the attributes in all the
relations of the schema are concatenated by a fixed order (e.g.,
lexicographical order).  In a similar manner, for the join conditions,
the encoding of all the join pairs are concatenated.

The selection conditions are specified on numerical/categorical
attributes.  For a range filter, $lb_j \leq A_j \leq ub_j$, on a
numerical attribute $A_j$, we normalize $lb_j$ and $ub_j$ to $[0, 1]$
by mapping $[lb_j, ub_j]$ to $[\frac{lb_j - min_j}{max_j - min_j},
  \frac{ub_j - min_j}{max_j - min_j}]$, where $[min_j, max_j]$ is the
domain of the attribute $A_j$. Thus, the representation is the
interval of two real values.
For an \IN~filter, $A_j~\IN~C$, on a categorical attribute $A_j$,
where $C$ is a subset of the attribute $A_j$'s domain $\{c_1, c_2,
\dots c_m\}$, a straightforward encoding is to build an $m$-dim bitmap
where its $k$-th bit is 1 if $c_k \in C$, otherwise 0.  This binary
representation is effective for attributes with small domain, however,
it is difficult to scale on a large domain where the predicate vector
is high-dimensional and sparse.  Therefore, for a large domain, we
adopt the factorized bitmap~\cite{DBLP:journals/corr/abs-2006-08109},
i.e., slicing the whole bitmap to chunks, and converting each chunk
into corresponding base-10 integer. Finally, the selection condition
is represented losslessly by $\lceil m / s \rceil$ integers, where $s$
is the length of the chunk.

Regarding join conditions, for each joinable attribute pair $(A_i,
A_j)$, we use a 3-bit bit-map to encode the join condition on this
pair, corresponding to the 3 comparison operators, $<, =, >$,
respectively, where `1' denotes there is a comparative condition on
$(A_i, A_j)$. For example, $A_i < A_j$, $A_i \geq A_j$, and $A_i \neq
A_j$ are encoded as `100', `011' and `101', respectively. The bit-map
`000' denotes that the query is free of join condition on the pair.
%
%

\section{An Overview}
\label{sec:ii}

\begin{figure}[t]
\centering
\includegraphics[width = 0.9\columnwidth]{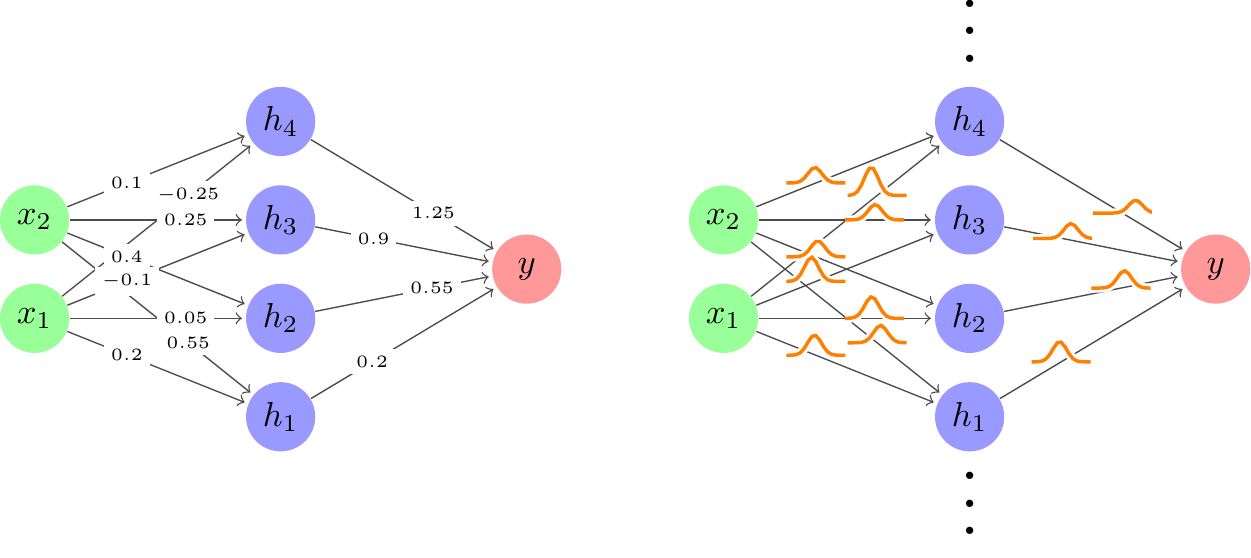}
\vspace*{-0.2cm}
\caption{Neural Network (left) vs. NNGP (right)}
\label{fig:nn-nngp}
\end{figure}

We present an NNGP overview for cardinality estimator that learns
a DL model using GP.
%
%
We discuss NNGP from the viewpoints of standard neural network and GP,
where NNGP exhibits the approximation capability of neural network and
can be solved by exact Bayesian inference as a regular GP.  Such NNGP
properties enable our cardinality estimator to be robust, lightweight,
and uncertainty-aware, compared with the existing DL-based estimators.
We show the differences between a neural network and NNGP in
Fig.~\ref{fig:nn-nngp}.

On the left, Fig.~\ref{fig:nn-nngp} shows a standard fully-connected
neural network, which is the building block of all the DL-based
estimators in Table~\ref{tbl:exist_approaches}.  The hidden layer is
a weighted linear transformation with nonlinearity that transforms input
representation to output in a layer by layer fashion.  Given an
empirical loss function, the parameters (i.e., the weights of the
linear transformation) fit to given training data (i.e., the
vectorized relational data or SQL query regarding cardinality
estimation) by forward-backward propagation algorithm.  As a
parametric model, the prediction on new input is determined by
the learned parameters.
%
%
Theoretically, neural network is able to approximate any given
continuous function~\cite{DBLP:journals/nn/HornikSW90,
  DBLP:journals/nn/Hornik91}, and achieve an arbitrary small
approximation error with infinite wide hidden layers.

On the right, Fig~\ref{fig:nn-nngp} shows NNGP, which is a special
class of Bayesian DL model, whose hidden layer has an infinite number of
neurons.
In statistical learning, Bayesian inference is a principled way to
describe prediction belief.  Bayesian DL is derived from developing
Bayesian inference on modern DL~\cite{DBLP:journals/tkde/WangY16}. Specifically, prior distributions are
imposed on the parameters of the neural network.  In other words,
given a set of training data, the posterior distribution of parameters
is inferred by Bayes rule.  And the prediction of a new input is a
probabilistic distribution computed by Bayesian model average that
ensembles all the models in the parameters space weighted by the
posterior of the parameters.\footnote{In general, exact analytical
  posterior distribution is intractable, thereby approximate inference
  such as variational inference, Markov chain Monte Carlo are
  adopted.}
NNGP, as a special type of Bayesian DL, is equivalent to GP in the
sense that any finite collection of outputs is a Gaussian
distribution, and its output is a summation of an infinite number of
i.i.d. random variables implied by Central Limit
Theorem~\cite{neal1996priors}.  The infinite hidden neurons are
composed of the set of GP basis functions, leading to a parametric,
non-stationary GP kernel.  Compared with regular GP with stationary
kernels, this DL-based kernel is more flexible to adapt to the
underlying data by exploiting the representation learning ability of
DL.  Our testing in \cref{sec:exp} shows that without DL, a simple GP
estimator fails to achieve an approaching or better performance,
compared with the DL cardinality estimators.

\comment{ On the right of Fig.~\ref{fig:nn-nngp} is a schematic
  illustration of Neural Network Gaussian Process (NNGP). In a nut
  shell, it distinguishes from the standard neural network in two
  aspects. First, the parameters are assumed to be i.i.d. random
  variables drawn from a prior distribution.  Second, its width of the
  hidden layers goes to infinity.  Imposing probability density on the
  parameters transforms the model into a Bayesian neural network.  The
  prediction of new input is a probabilistic distribution computed by
  Bayesian model average that ensembles all the models in the
  parameters space weighted by the posterior probability of the
  parameters.  The infinite number of hidden units pushes this
  Bayesian neural network to become a special GP whose infinite number
  of basis functions are specified by the hidden units of neural
  network.  }


From the perspective of Bayesian DL, NNGP inherits the advantages of
Bayesian approach while keeping the universal approximation of neural
network.  With the parameter prior, NNGP converts from neural network
parameter learning to the prior's hyper-parameter learning so that it
overcomes the over-parametrization of neural network.  Instead of
betting on one parameter configuration, NNGP ensembles an infinite number
of plausible neural network models in the space of a given
architecture and prior family, thereby provides a robust
approximation.  The predictive distribution naturally conveys the
prediction uncertainty regarding the model's posterior and indicates
out-of-distribution testing points.

From the perspective of GP, NNGP is a nonparametric model with the
infinite number of basis functions.
Here, nonparametrics does not mean the model is parameter-free,
but in the sense that the model can not be specified in a finite number
of parameters.  As a nonparametric model, NNGP utilizes a much larger
model space to model distribution-free data. In other words, it does
not assume data to be modeled is i.i.d. or generated from a specified
distribution as what DL models require~\cite{lee2004bayesian}.  The
property of distribution-free promotes deploying database-oriented
learning tasks (e.g., cardinality estimation), since real data and
queries in DBMS are too large and diverse. It is difficult to assume
that they are subject to some distribution where both data and queries
may evolve over time.  Therefore, compared to general DL models, NNGP
does not require large volume training samples to approximate an
i.i.d. assumption.
In addition, the learning paradigm of NNGP converts from parameter
learning for DL to kernel learning for GP, which avoids the
approximation inference of Bayesian DL by manipulating the infinite
wide neural network implicitly.  More concretely, NNGP under certain
neural network configuration (e.g., $\relu$ nonlinearity) has an
analytical kernel function, which means training and prediction of
this special type of DL model can be solved by pure statistical method
in closed-form (i.e., exact Bayesian inference) as for a regular GP.
In our experiments of \cref{sec:exp}, we verify training an NNGP
estimator on the fly only consumes several seconds, up to 1-2 orders
faster than corresponding DL estimators.
%

\begin{figure}[t]
\centering
\begin{tabular}[t]{c}
\subfigure[\forest]{
\label{fig:visual:kernel:forest}
 \includegraphics[width=0.42\columnwidth]{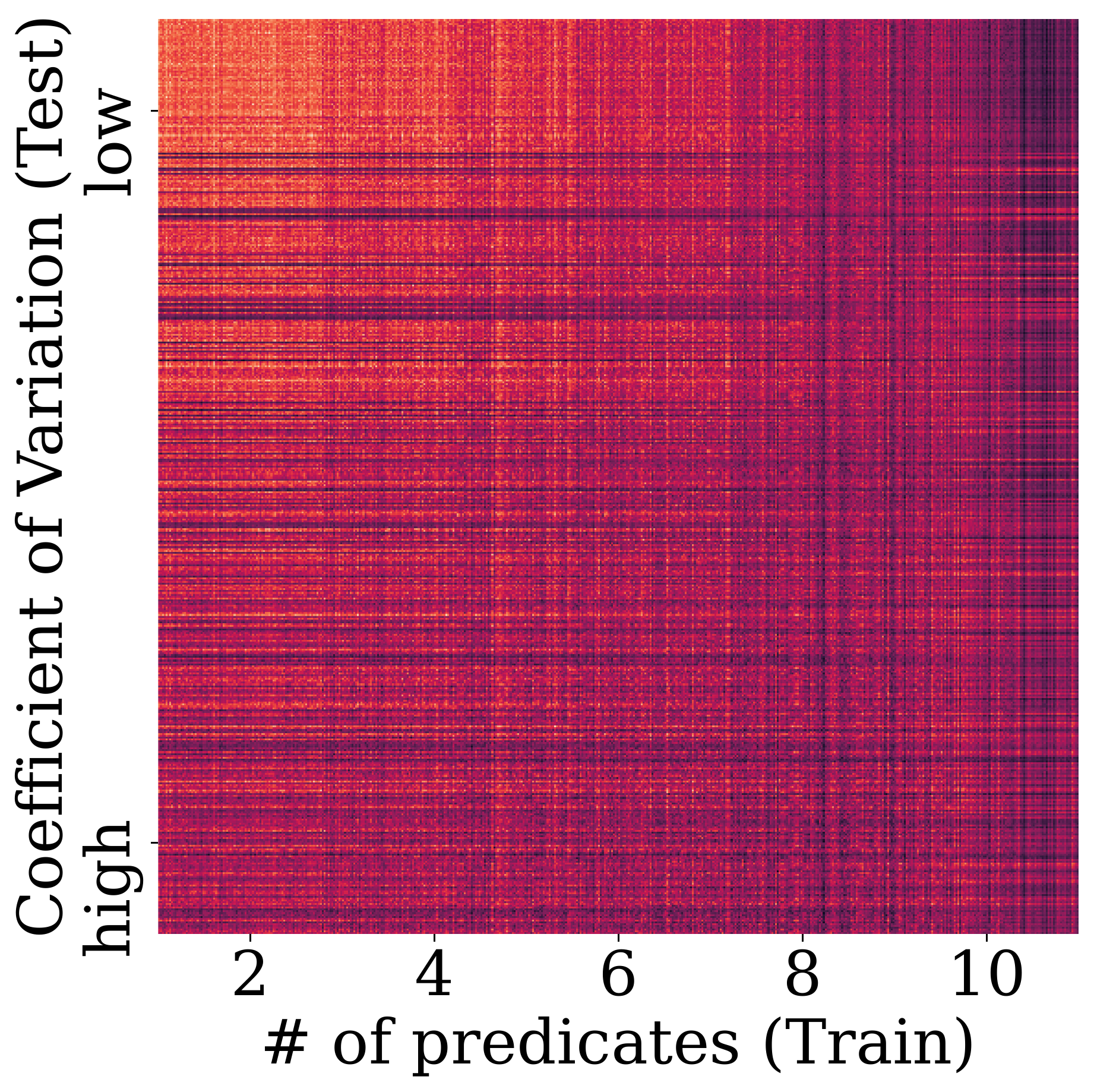}
}
\subfigure[\tpcds]{
\label{fig:visual:kernel:tpcds}
 \includegraphics[width=0.42\columnwidth]{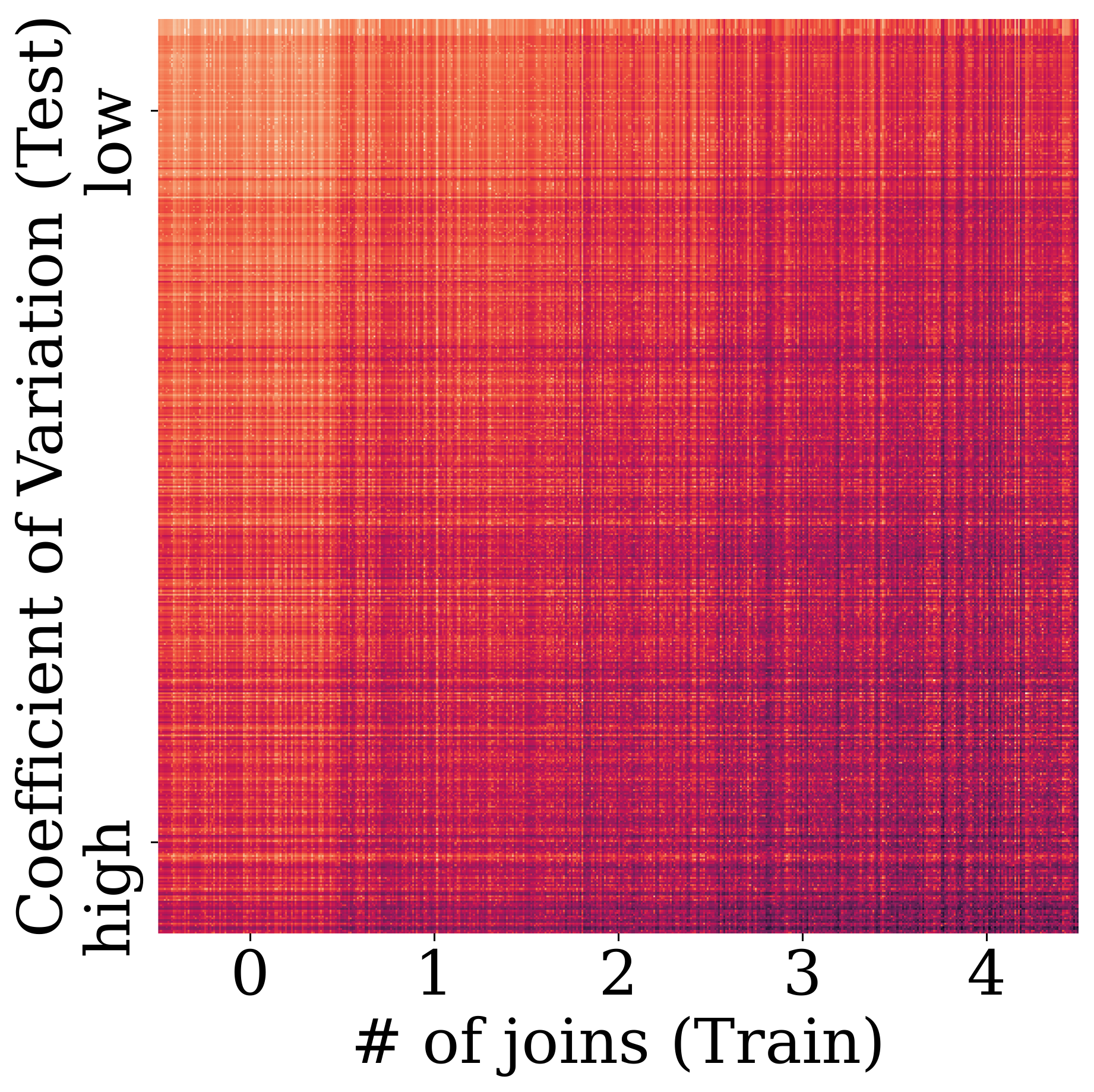}
}
\end{tabular}
\vspace*{-0.2cm}
\caption{Test-Train Kernels of NNGP}
\label{fig:visual:kernel}
\vspace*{-0.0cm}
\end{figure}

We show NNGP, as a kernel method, from the point of neural network
enhanced kernel function in Fig.~\ref{fig:visual:kernel}. In
Fig.~\ref{fig:visual:kernel}, we visualize two kernel matrices of NNGP
estimator by taking vectorized SQL queries as the input, which are
created by $500$ training queries and $500$ test queries.  The queries
for Fig.~\ref{fig:visual:kernel:forest} are on a single relation
\forest, with $2 \sim 10$ selection conditions, while the queries for
Fig.~\ref{fig:visual:kernel:tpcds} are join queries over $0 \sim 5$ relations
in \tpcds. The kernel matrices are ordered by the number of
selection/join conditions for the training query and the coefficient of variation of
the predictive distribution for the test query.  In
Fig.~\ref{fig:visual:kernel}, the lighter the color, the larger the
inner product of the infinite hidden representations of a train-test
query pair, which indicates the larger the similarity between the pair.
There are two key observations.  First, the uncertainty of the
prediction of a test query is highly correlated with its similarity to
the training queries. The lower the uncertainty, the larger the
similarity to all the 500 training queries.  This observation supports
our intuition.  Second, a query with more join conditions tends to be
more dissimilar to other queries.  The NNGP kernel provides a simple
yet effective mechanism to compare input similarity regarding
transformed infinite feature space. The knowledge of the training data
is persisted in the kernel matrix to smooth the prediction, with
similarity measures between training/testing data. 

We discuss NNGP with the existing DL approaches for cardinality
estimation following a recent experimental study
in~\cite{DBLP:journals/corr/abs-2012-06743} that analyzes and compares
5 learned cardinality estimators over single relation.
In~\cite{DBLP:journals/corr/abs-2012-06743}, the authors identify a
set of behavior logics, as the inductive biases the learned estimators
are expected to capture, namely, consistency, monotonicity, stability,
and fidelity.
All the query-driven DL estimators only preserve the stability as they
model the estimation as a regression task. The data-driven DL
estimator \NeuroCard supports fidelity but cannot satisfy
others, whereas the data-driven DL estimator \DeepDB satisfies all the
logics since it relies on hierarchical density factorization.
\comment{
Fortunately, a good news is that the illogical deviation of ML/DL does
not impedes the learned estimators to deliver more accurate prediction
than traditional methods.
Meanwhile, the study also reveals some
problems and challenges these learned estimators face, including the
longer training time, the extra cost for hyper-parameters tuning on DL
based estimators, the effectiveness and efficiency of model adaptation
in dynamic environments, trustworthy and interpretable prediction.
}
Our NNGP estimator is a supervised learning based estimator. Thereby
NNGP only supports stability like all the other query-driven
regressors. However, the NNGP estimator distinguishes from the
existing learned estimators from its small tuning cost and swift model
construction, which are promising for fast adaption in dynamic
environments.  NNGP supports Bayesian based uncertainty quantification
regarding the acquired knowledge, and kernel learning enables
kernel based feature understanding and
selection~\cite{DBLP:journals/jmlr/SongSGBB12,
  DBLP:conf/nips/ChenSWJ17}.

\comment{
Neural network provides a highly flexible, parametric basis function, which is optimized to adapt to the data, aiming to project the input data into useful feature space. Infinite wide neural network are proved to be able to approximate any given continuous function , i.e., Universal Approximation Theorem.
Despite the powerful approximation, neural network has its drawbacks. 
First, well-performed neural networks relies on some   assumptions regarding the underlying distribution, e.g., data are i.i.d.. Thereby, in the real application, it needs training over large amount of data to reduce variance, pursuing these assumptions are approximately true. 
Second, betting on one parameter configuration, neural networks are hardly to convey their prediction belief, the correctness likelihood of their predictions. 
As moving away from the observed data, neural networks often mis-calibrate in the sense that the predictions are typically overconfident~\cite{DBLP:conf/icml/GuoPSW17}.
This is also one reason why neural networks lack interpretability. 
Third, in practice, it is infeasible to represent and manipulate infinite wide neural networks within limited computing resources.

Incorporating GP deals with these drawbacks. NNGP does not make the i.i.d. assumption over the data distribution. The model itself, i.e., the set of parameter is a random variable, and is determined by the posterior distribution given the observed data. 
Uncertainty of prediction is measured by the uncertainty of the model's posterior.   
Therefore, the model is robust to unseen data and requiring relatively less training data. The NNGP kernel has many fewer parameters to be optimized, i.e., the type of the prior distributions and their mean and variances (Eq.~(\ref{eq:nngp:prior})), corresponding to the hyper-parameters of the classical machine learning. 
Moreover, the kernel method enables operating infinite dimensional feature space implicitly. 
Specifically, the kernel matrix persists the inner product of the representation derived from the infinite dimensional basis function as Eq.~(\ref{eq:rhbs:kernel}):
\begin{align}
\Cov_{x_1, x_2} = \langle [h_1(x_1), h_2(x_1), \cdots ] {,} [ h_1(x_2), h_2(x_2), \cdots ]\rangle
\label{eq:rhbs:kernel}
\end{align}
It bypasses the explicit computation of the feature representation. 
}

Finally, the way we use GP is different from GP used for database
configuration tuning~\cite{DBLP:conf/sigmod/AkenPGZ17,
  DBLP:journals/pvldb/DuanTB09}.  First,
in~\cite{DBLP:conf/sigmod/AkenPGZ17, DBLP:journals/pvldb/DuanTB09}, GP
is used to model the concerned performance given a set of input
configurations, and Bayesian optimization is used to search for a 
better one among a set of configurations.
In our approach, 
%
%
%
%
the GP regressor is
the cardinality estimator.  Second, the GP used in
\cite{DBLP:conf/sigmod/AkenPGZ17, DBLP:journals/pvldb/DuanTB09} is
equipped with stationary kernel functions.  For cardinality estimation
with a larger feature space (i.e., the query space and a higher
requirement of accurate prediction), stationary kernel functions are
not flexible in terms of adapting to the input data.  We use
non-stationary kernel functions that belong to a particular type
of neural network.
%
%

\section{NNGP: from NN to GP}
\label{sec:bg}

In this section, we discuss Bayesian learning, and its difference from
standard learning, introduce GP and present NNGP. 

\subsection{Bayesian Learning}
In standard parametric ML/DL, a model to learn is a function of input
$x$, $f(x, w)$, parameterized by $w$. Learning is
to optimize a specified loss function $\loss(f(x, w), y)$ to
fit the parameter $w$ to training data $(x, y)$ in a
training set $(X, Y)$.  In contrast to learning by
\emph{optimization}, Bayesian learning is to learn by
\emph{marginalization}. The parameter $w$ is assumed to be a random
variable drawn from a prior distribution $p(w)$. Given the observed
training data $(X, Y)$, the posterior distribution of $w$
can be inferred by Bayes rule (Eq.~(\ref{eq:bnn:bayes})).
\begin{align}
    \label{eq:bnn:bayes}
    p(w| Y, X )  = \frac{ p(Y | X, w) p(w) }{ p(Y | X)} = \frac{ p( Y | X, w) p(w) } 
 									  { \int p(Y | X, w) p(w) dw} 
\end{align} 
To infer the target value for testing data $X^*$, the
predictive probability $p(Y^{*}| X^{*}, Y, X)$ is
computed by applying the probabilistic sum and product rules, assuming the training and
testing are conditional independent regarding $w$.
\begin{align}
   p(Y^{*}| X^{*}, Y, X) 
   = \int p(Y^{*}| X^{*}, w) p( w | Y, X) d w \label{eq:ensemble:bayes}
\end{align}
The predictive distribution of Eq.~(\ref{eq:ensemble:bayes})
represents \emph{Bayesian model
  average}~\cite{DBLP:conf/nips/WilsonI20}. That is, instead of
relying on a single prediction of one model with a single
configuration of parameters, Eq.~(\ref{eq:ensemble:bayes}) ensembles
all the models with all possible configurations of the parameters
${w}$, weighted by the posterior of the parameters, $p({w}|
{Y}, {X})$, from Eq.~(\ref{eq:bnn:bayes}), by marginalization of
$w$.  Therefore, the predictive distribution does not depend on any
specific parameter configuration.  In contrast, classical training of
parametric model aims to find one configuration $\hat{{w}}$ that
maximizes the likelihood of the observed data or minimizes the
empirical loss in equivalence.  In other words, the posterior
distribution $p({w} | {Y}, {X}) = 1$ for ${w} =
\hat{{w}}$ and $0$ otherwise, leading to the model inference as
Eq.~(\ref{eq:ensemble:classical}).
\begin{align}
\label{eq:ensemble:classical}
       p(Y^{*}| {X}^{*}, Y, {X}) = p(Y^{*}|{X}^{*}, \hat{{w}}), \hat{{w}} = \mathop{\arg\max}\limits_{{w}} p(Y | {X}, {w})
\end{align}
Comparing the inference of Eq.~(\ref{eq:ensemble:bayes}) and
Eq.~(\ref{eq:ensemble:classical}), if the weights posterior $p({w}|
{Y}, {X})$ has a flat distribution and $p({Y}^{*}|{X}^{*},
\hat{{w}})$ varies significantly where the posterior has mass, the
discrepancy of prediction by Bayesian model average and classical
approach tends to be large.  One well-known reason is the observed
data is insufficient or deviating from the features of the test data, where
the observed data cannot well infer the weight posterior in principle.
The Bayesian learning calibrates this kind of uncertainty, a.k.a.,
\emph{epistemic uncertainty}.

\begin{figure}[t]
\centering
\scriptsize{
\begin{tabular}[t]{p{3cm}<{\centering} p{3cm}<{\centering}} 
 $f(x) = \bm{w} x + b$ & \makecell[c]{ $f(x) = \bm{w} x + b$ \\ $\bm{w} \sim \Norm(0, \sigma^2_w)$, $b \sim \Norm(0, \sigma^2_b)$} \\ 
\end{tabular}
}
\vfill
\begin{tabular}[t]{c}
\subfigure[Linear Regression]{
\label{fig:example:lr-blr:lr}
 \includegraphics[width=0.42\columnwidth]{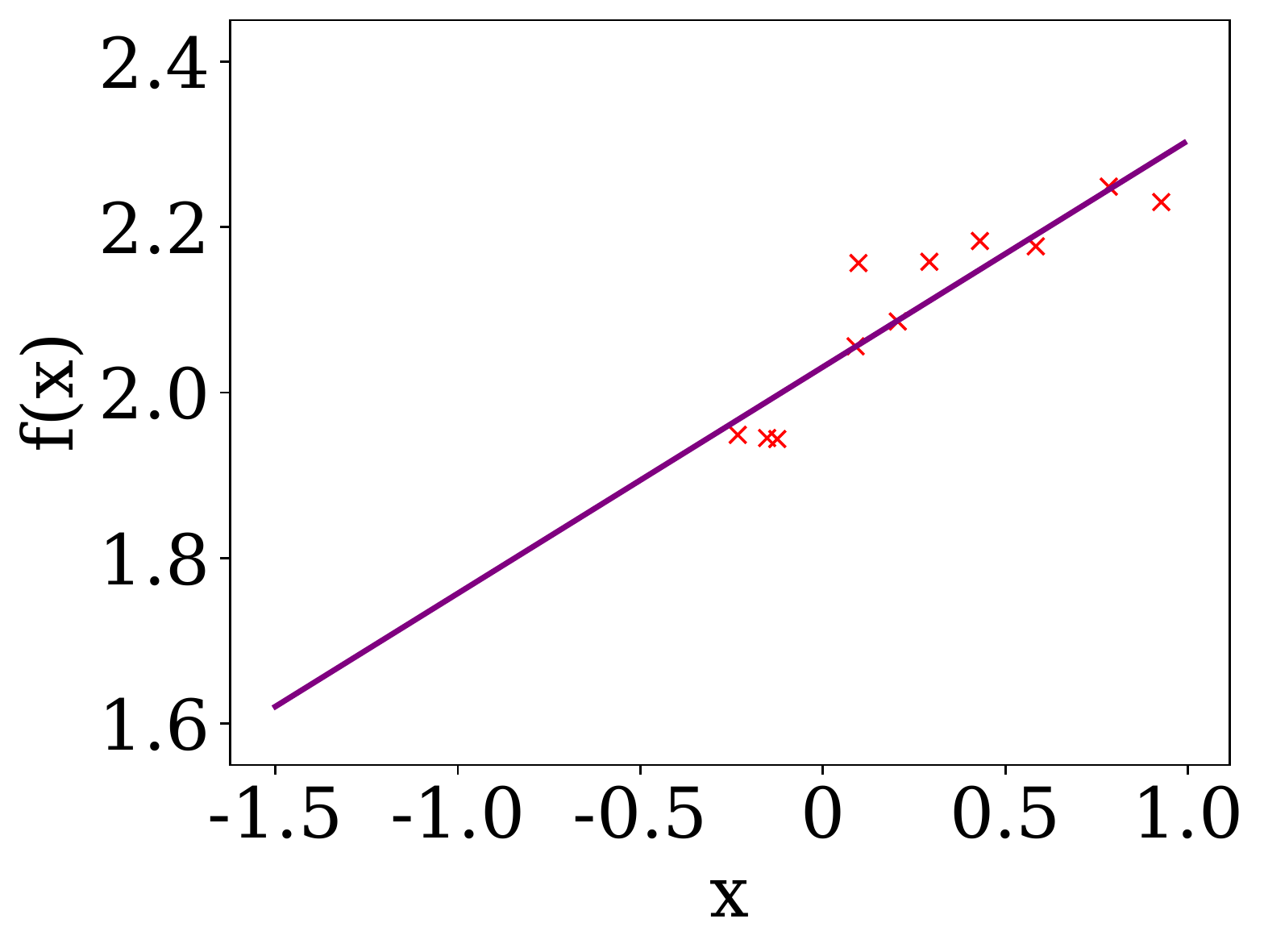}
}
\subfigure[Bayesian Linear Regression]{
\label{fig:example:lr-blr:blr}
 \includegraphics[width=0.42\columnwidth]{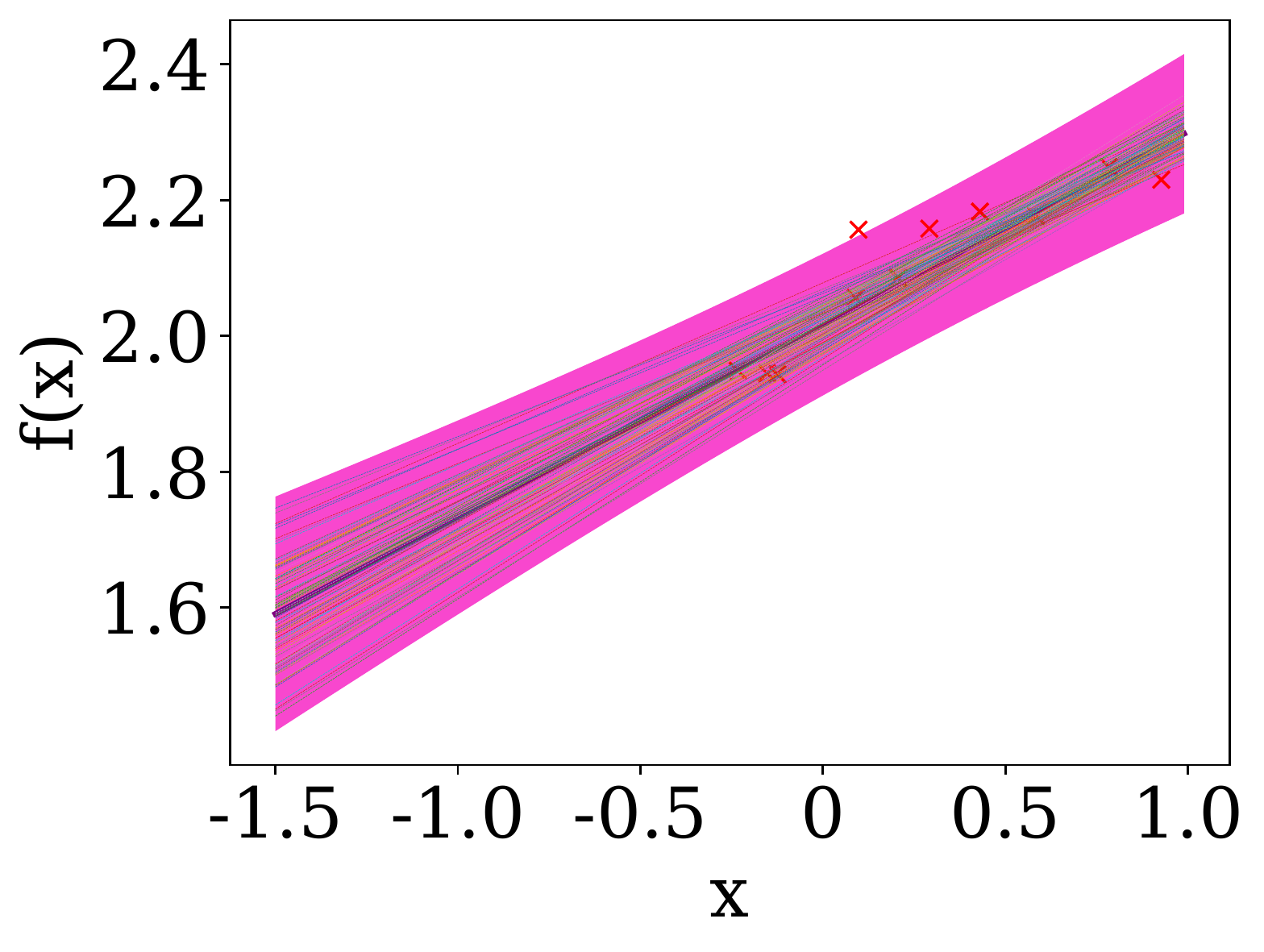}
}
\end{tabular}
\vspace*{-0.3cm}
\caption{LR vs. BLR}
\label{fig:example:lr-blr}
\vspace*{-0.1cm}
\end{figure}

\begin{example} A simple example of Linear Regression (LR) and
  Bayesian Linear Regression (BLR) is shown in
  Fig.~\ref{fig:example:lr-blr} to illustrate the differences between
  standard learning and Bayesian learning.  
In LR (Fig.~\ref{fig:example:lr-blr:lr}), a linear function $f(x) = wx
+ b$ is used to fit the training data points, minimizing the square
error. The parameters $w$ and $b$ can be solved analytically by least
squares or gradient descent.  In BLR
(Fig.~\ref{fig:example:lr-blr:blr}), it does not aim to solve a
deterministic value for $w, b$, instead the posterior distributions of
$w$, $b$ are computed, given the training data, assuming the priors of
$w, b$ are simple Gaussian distributions with zero means.  The model
is no longer a linear function but a random variable determined by its
parameter posterior, which forms a function space with infinite linear
functions. The color lines in Fig.~\ref{fig:example:lr-blr:blr}
indicate some samples in the function space.  The derived predictive
distribution (Eq.~(\ref{eq:ensemble:bayes})) of BLR is a Gaussian
distribution, thereby the shaded area in
Fig.~\ref{fig:example:lr-blr:blr} delineates the 95\%-confidential
interval.  We can observe that in the range with few training data,
i.e., $x \in [-1.5, -0.5]$, the predictive uncertainty tends to be
large, reflect to the fanout of the sampled functions and wider
confidential interval. BLR only has two hyper-parameters to be set,
which are the variances $\sigma^2_{w}$, $\sigma^2_{b}$ of the
parameter prior.  The hyper-parameters can be tuned by Bayesian model
selection~\cite{DBLP:books/lib/RasmussenW06}.
\end{example}

\subsection{Gaussian Process}

Given a set of $N$ training data points ${X} = \{{x}_1,\cdots,
{x}_N \}$ and a model $f({x})$, GP is used in modeling the joint
distribution of the model's predictions $f({X}) =
\{f({x}_1),\cdots, f({x}_N) \}$.  Consider a simple linear model
in Eq.~(\ref{eq:linear:basis}), with fixed \emph{basis function} $\phi(\cdot)
= \{\phi_1, \cdots, \phi_m\}$, (e.g., polynomial, radial basis
function), which is a projection from original input to a feature
space.
\begin{align}
\label{eq:linear:basis}
    f({X}) = \sum_{j=1}^{m} w_j \phi_j({X}) = \bm{\Phi} {w}
\end{align}
In Eq.~(\ref{eq:linear:basis}), $\bm{\Phi} \in \Real^{N \times m}$ is
the design matrix where $\bm{\Phi}_{nj} = \phi_j({x}_n)$, i.e., the
value of the $j$-th basis function for the $n$-th data point.  Suppose
a zero mean Gaussian prior is put on the weights ${w}$, where
$\Ident \in \Real^{N \times N}$ is the Identity matrix.
\begin{align}
\label{eq:gp:prior}
    {w} \sim \Norm(\Zero, \sigma_{w}^2\Ident)
\end{align}
Since $f({X})$ is a linear transformation of ${w}$, $f({X})$
is also a Gaussian distribution as Eq.~(\ref{eq:gp:joint:dist}), where
${\epsilon} \sim \Norm(\Zero, \sigma_{\epsilon}^2 \Ident)$ is an extra noise on $f({X})$.
\begin{align}
\label{eq:gp:joint:dist}
    f({X}) \sim \Norm(\Zero, \sigma_{w}^2 \bm{\Phi} \bm{\Phi}^{T} + \sigma_{\epsilon}^2\Ident)
\end{align}
Here, Eq.~(\ref{eq:gp:joint:dist}) shows the function $f({x})$, as
a random variable, is a \emph{Gaussian process} by definition such
that for any finite selection of points ${X}$, $f({X})$ is a
joint Gaussian distribution~\cite{mackay1998introduction}.
The covariance matrix $\Cov_{X,X} = \sigma_{w}^2 \bm{\Phi} \bm{\Phi}^{T} +
\sigma_{\epsilon}^2\Ident$, a.k.a., the \emph{kernel matrix}, models
the similarity of data points. Specifically, the entry $\Cov_{ik} =
\sigma_{w}^2 \langle \phi({x}_i),  \phi({x}_k) \rangle +
\sigma_{\epsilon}^2\Ident_{ik}$ measures the similarly of points
${x}_i, {x}_k$ under basis function $\phi(\cdot)$.  A simplest
example is that $\phi(\cdot)$ is the identity function, i.e.,
$\phi({x}) = {x}$, its model is equivalent to BLR. 

\comment{ 
To make prediction for testing data points ${X}^* = \{x^{*}_1, \cdots, x^{*}_M \}$, 
we need to compute the conditional
distribution in Eq.~(\ref{eq:prob:cond}).
\begin{align}
\label{eq:prob:cond} 
p(f({X}^{*}) | f({X}) ) = \frac{p(f({X}^{*}),  f({X}) )}{p(f({X}))}
\end{align}
where $p(f({X}), f({X}^{*}))$ of Eq.~(\ref{eq:prob:cond}) is a
joint Gaussian distribution as $f({x})$ is a GP,
as shown in Eq.~(\ref{eq:joint:gaussian}).  Here, the covariance matrix
is a $(N + M) \times (N + M )$ dimensional matrix, which can be
blocked into the ${X}, {X}$ kernel, ${X}, {X}^{*}$ kernel
and ${X}^{*}, {X}^{*}$ kernel matrices.
\begin{align}
\label{eq:joint:gaussian}
\begin{bmatrix}
f({X}) \\
f({X}^{*})
\end{bmatrix}\sim \Norm\left(\Zero,\begin{bmatrix}
\Cov_{X,X} & \Cov_{X, X^{*}} \\
\Cov_{X, X^{*}}^{T} & \Cov_{X^{*}, X^{*}}
\end{bmatrix}\right)
\end{align}
Take advantage of the property of Gaussian distribution: if two sets
of variables are jointly Gaussian, then the conditional distribution
of one set conditioned on the other is still a Gaussian
distribution. Gaussian can be solved analytically as
follows~\cite{DBLP:books/lib/Bishop07}.
}
To make prediction for testing data points ${X}^* = \{x^{*}_1, \cdots, x^{*}_M \}$, 
we need to compute the conditional distribution $p(f({X}^{*}) | f({X}) )$ as the prediction. 
It is also proved to be a Gaussian distribution as Eq.~(\ref{eq:gp:pred}), 
where $\Cov_{X, X} \in \Real^{N \times N}$, 
$\Cov_{X, X^*} \in \Real^{N \times M}$, $\Cov_{X^*, X^*}\in \Real^{M \times M}$ 
are the ${X}, {X}$ kernel, ${X}, {X}^{*}$ kernel
and ${X}^{*}, {X}^{*}$ kernel matrices, respectively. 
\begin{align}
    f({X}^{*}) &| f({X}) \sim \Norm({\mu}, {C}) \label{eq:gp:pred}\\
    {\mu} &= \Cov_{X, X^{*}}^T\Cov_{X, X}^{-1}f({X}) \label{eq:gp:mean} \\
    {C} &= \Cov_{X^*, X^*} - \Cov_{X, X^{*}}^T\Cov_{X, X}^{-1}\Cov_{X, X^{*}} \label{eq:gp:cov}
\end{align}
Given the ground truth of the training data ${X}$, denoted as $Y$, and the prediction target for ${X}^{*}$ is ${Y}^*$, the expectation of the prediction  is $\Expect[Y^{*}] = \Cov_{X, X^{*}}^T\Cov_{X, X}^{-1}{Y}$, derived from Eq.~(\ref{eq:gp:mean}). 
Assume a matrix of functions ${h}({X}^{*}) = \Cov_{X, X^{*}}^T\Cov_{X, X}^{-1}$, we have  $\Expect[ f({X}^{*})]  = {h}({X}^{*}) Y$, indicating GP regression is a weighted linear smoother over the observed target value $Y$. 
The weight function ${h}$ is determined by the train-train and train-test kernels. 
Meanwhile, the diagonal element of matrix ${C}$ in Eq.~(\ref{eq:gp:mean}) measures the variance of the prediction.
With this predictive Gaussian distribution, we can easily compute the
$\delta$-confidential interval of $f({x}^{*})$
as
$    [\mu_{x^*} - {q}_{\delta} diag(C)_{x^*}, \mu_{x^*} + {q}_{\delta} diag(C)_{x^*}]$,
where $q_{\delta}$ is the $\delta$-quantile of $\Norm(0, 1)$.
Intuitively, the expectation $\Expect[{y}^{*}]$ should be treated
as the explicit prediction $\hat{y}$.  In practical applications, as
it aims to minimize the predictive loss given an empirical loss
function, the prediction is to find $\hat{y}$ that minimize the
\emph{expected loss} by averaging the empirical loss
$\mathcal{L}_{\kw{exp}}$ w.r.t. the predictive distribution as
Eq.~(\ref{eq:gp:exploss}).
\begin{align}
\mathcal{L}_{\kw{exp}} = \int \mathcal{L}_{\kw{emp}}({y}^{*}, {\hat{y}}) p(y^* | x^*,  Y, X) dy^{*}
\label{eq:gp:exploss}
\end{align}

\comment{
\begin{figure}[t]
\centering
\includegraphics[width = 0.8\columnwidth]{fig/gp_example.pdf}
\caption{$1$-dim Gaussian Process with RBF Kernel}
\label{fig:gp:example}
\end{figure}

\begin{example}{A $1$-dim Gaussian Process with RBF Kernel. Fig~\ref{fig:gp:example} shows a Gaussian Process trained by 7 data points. 
The data are generate from $y = \sin(x) + c \sigma$ where $x = \{ -3, -2, -1, 0, 1, 2, 3\}$, $\sigma$ is a random noise from $\Norm(0, 1)$ and $c$ is a noise coefficient set to 0.4. The kernel function is the stationary RBF Kernel (Eq.~(\ref{eq:kernel:rbf})) and the hyper-parameter length scale $l$ is set to 1.0.
\begin{align}
\label{eq:kernel:rbf}
    \Cov_{RBF} = \exp \left(\frac{|| x_i - x_j ||^2}{2l^2}\right)    
\end{align}
}

In Fig.~\ref{fig:gp:example}, the shade area delineates the 95\% confidential interval of the prediction on the test data. We can observe that as the test data is far away from the training data, the uncertainty becomes large.  
The 3 dashed lines denotes 3 samples of $f(x)$ drawn from the posterior distribution, respectively. 
The final prediction is a Bayesian model average that ensembles all the infinite sampled model. 
\end{example}

The predictive distribution highly depends on the kernel matrix
$\Cov$.  For a fixed basis function, the kernel matrix is
parameterized by the parameters $\theta = \{ \sigma_{w},
\sigma_{\epsilon}\}$, where $\sigma_{w}$ delineates the length scale
of the linear correlation and $\sigma_{\epsilon}$ delineates the
precision of the noise, respectively.  These parameters correspond to
the hyper-parameters of in a standard parametric model, which can be
optimized to adapt to the training data.  The optimization is
conducted by maximized the (log)likelihood of the observed data
$(\bm{X}, \bm{y})$,
\begin{align}
\label{eq:gp:likelihood}
\mathcal{L}=\log p(\bm{y} | \bm{X}, \theta) = -\frac{1}{2}  |\Cov_{X, X}| -\frac{1}{2} \bm{y}^T \Cov_{X, X}^{-1} \bm{y} - \frac{N}{2} \log{2\pi}
\end{align}
There is no analytical solution for maximizing the likelihood of Eq.~(\ref{eq:gp:likelihood}) but the objective is differentiable. We can use gradient decent to minimize the negative of $\mathcal{L}$ via computing the gradients. 
} 

\comment{
We distinguish of Gaussian process from standard machine learning models in the perspective of Bayesian inference. 
Our model $f(\bm{x})$ itself is a random variable as prior knowledge is incorporated on the weights $\bm{w}$ (Eq.~(\ref{eq:gp:prior})).
Given the weights $\bm{w}$,  the data likelihood is also a Gaussian 
\begin{align}
\label{eq:likelihood}
    p(\bm{y} | \bm{X}, \bm{w}) = \Norm(\bm{\Phi}\bm{w}, \sigma_{\epsilon}^2 \Ident)
\end{align}
Via Bayes rule, the posterior probability of $\bm{w}$ is a Gaussian distribution in Eq.~(\ref{eq:rule:bayes}), which indicate the posterior belief of the weights given the observed data $(\bm{X}, \bm{y})$.  
\begin{align}
    \label{eq:rule:bayes}
    p(\bm{w}| \bm{y}, \bm{X} ) \propto p(\bm{y} | \bm{X}, \bm{w}) p(\bm{w}) = \Norm(\bm{\mu}', \bm{C}') 
\end{align} 
where
\begin{align}
    \label{eq:gp:posterior}
    \bm{C}' = (\sigma_{\epsilon}^{-2} \bm{\Phi}^{T} \bm{\Phi} + \sigma_{w}^{-2} \Ident) ^{-1}, 
    \bm{\mu}' = \sigma_{\epsilon}^{-2} \bm{C}' \bm{\Phi}^{T} \bm{y}
\end{align}

To predict the target value for test data $\bm{X}^*$, we compute the probability $p(\bm{y}^{*}| \bm{X}^{*}, \bm{y}, \bm{X})$ by applying the sum and product rules of probability by Eq.~(\ref{eq:ensemble:bayes}). Since the likelihood (Eq.~(\ref{eq:likelihood})) and posterior (Eq.~(\ref{eq:gp:posterior})) are both Gaussian,  the prediction is a Gaussian distribution in Eq.~(\ref{eq:pred:dist}). This predictive distribution is equivalent to that of Eq.~(\ref{eq:gp:pred}) after some transformations~\cite{DBLP:books/lib/RasmussenW06}. 
\begin{align}
\label{eq:ensemble:bayes}
   p(\bm{y}^{*}| \bm{X}^{*}, \bm{y}, \bm{X}) &= \int p(\bm{y}^{*}|\bm{X}^{*}, \bm{w}) p(\bm{w}| \bm{y}, \bm{X}) d\bm{w} \\
&= \Norm({\phi}(\bm{X}^{*}) \bm{\mu}', {\phi}(\bm{X}^*)\bm{C}' {\phi}^{T}(\bm{X}^*)) \label{eq:pred:dist}
\end{align}
The inference of Eq.~(\ref{eq:ensemble:bayes}) represents \emph{Bayesian model average}~\cite{DBLP:conf/nips/WilsonI20}, i.e., instead of relying on a single prediction of one model with a single configuration of parameters, Eq.~(\ref{eq:ensemble:bayes}) ensembles all the models with all possible configurations of the parameters $\bm{w}$, weighted by the posterior of the parameters, by marginalization of $w$.  Therefore, the predictive  distribution does not depend on any specific parameter configuration. 
In contrast, classical training of parametric model aims to find one configuration $\hat{\bm{w}}$ that maximizes the likelihood of the observed data. In other words, the posterior distribution $p(\bm{w} | \bm{y}, \bm{X}) = 1$ for $\bm{w} = \hat{\bm{w}}$ and $0$ otherwise, leading to the model inference as Eq.~(\ref{eq:ensemble:classical}).  
\begin{align}
\label{eq:ensemble:classical}
       p(\bm{y}^{*}| \bm{X}^{*}, \bm{y}, \bm{X}) = p(\bm{y}^{*}|\bm{X}^{*}, \hat{\bm{w}}), \hat{\bm{w}} = \mathop{\arg\max}\limits_{\bm{w}} p(\bm{y} | \bm{X}, \bm{w})
\end{align}
Comparing the inference of Eq.~(\ref{eq:ensemble:bayes}) and Eq.~(\ref{eq:ensemble:classical}), if the weights posterior $p(\bm{w}| \bm{y}, \bm{X})$  has a flat distribution and $p(\bm{y}^{*}|\bm{X}^{*}, \hat{\bm{w}})$ varies significantly where the posterior has mass, the discrepancy of prediction by Bayesian model average and classical approach tends to be large. 
One common reason for this situation is the observed data is insufficient or deviating the features of the test data, where  the observed data cannot well infer the weight posterior in principle. 
The Bayesian learning calibrates this kind of uncertainty, a.k.a., \emph{epistemic uncertainty}. In the view of kernel learning, similarity of data points are leveraged to derive the joint distribution and conditional distribution of the target value. In the view of weight space, the model is an ensemble of infinite models in the overall weight space.  
}

\subsection{Neural Network Gaussian Process}

GP is a stochastic process with a fixed basis function. If the
basis is fixed, the model is linear w.r.t. the parameters, and the
kernel function $\Cov$ as well as the predictive distribution
(Eq.~(\ref{eq:gp:pred})) are analytically tractable.  The limitation
of a fixed basis function is its incapability of adapting to the training
data. In general, a model with an adaptive basis function (e.g., neural
networks) can be a potential extension, but it is much difficult to
treat it analytically~\cite{DBLP:books/lib/RasmussenW06}.
To address it, the authors in~\cite{neal1996priors} show that there is
a special case where the neural network has an infinite number of hidden
units. With such findings, some complex neural network architectures
with infinite wide hidden layers are proved to be GP
such as convolutional neural
network~\cite{DBLP:conf/iclr/NovakXBLYHAPS19,
  DBLP:conf/iclr/Garriga-AlonsoR19}, recurrent neural
network~\cite{DBLP:conf/nips/Yang19},
attention~\cite{DBLP:conf/icml/HronBSN20} and graph neural
network~\cite{DBLP:journals/corr/abs-2002-12168}.

We discuss the foundation, the infinite wide multilayer perceptrons
which we use for cardinality estimation. We explain it using a single
hidden layer feed forward neural network, $f({x})$,
which takes $d$-dim vector ${x} = [x_j]^{d}$ as input, and predicts a scalar
value $y$. Here, $\act(\cdot)$ is
the nonlinear activation function, ${b}^0 =[b^0_j]^{m}$, $b \in
\Real$ is the bias term, and both ${w}^0 = [w^0_{ij}]^{m \times d}$
and ${w} = [w_i] ^{m}$ are the weights of the hidden layer and the
output layer,
respectively. Eq.~(\ref{eq:nn:postact})-(\ref{eq:nn:output}) show the
computation on each neuron of the hidden and output layers, where
$h_j$ in Eq.~(\ref{eq:nn:postact}) is the post-activation of the
$j$-th hidden unit.
\begin{align}
    h_i({x}) &= \act( \sum_{j=1}^d w^{0}_{ij} {x}_j + b^0_j)  \label{eq:nn:postact} \\
    f({x}) &= \sum_{i = 1}^{m} w_i h_i({x}) + b \label{eq:nn:output}   
\end{align}
In classical machine learning, it optimizes the parameters
${w}$ and ${w}^0$ directly, under a specified loss function
as an objective, by back propagation algorithm.
Note that, even though the neural network is nonlinear, it can be
regarded as a linear combination of a collection of parametric basis
functions $\{h_1({x}) \cdots, h_m({x})\}$ in
Eq.~(\ref{eq:nn:output})~\cite{DBLP:books/lib/Bishop07}.  The basis
functions are parametrized by the weight ${w}^0$ and will be
trained to adapt to the training data.
Under the assumption that, for each layer, the weight and bias element
parameters have i.i.d. prior densities, we have
\begin{align}
    \label{eq:nngp:prior}
  w^{0}_{ij} &\sim \Dist(0, \sigma_{w}^2/d), b^{0}_{j} \sim \Dist(0, \sigma_{b}^2) \\ 
   w_{i}  &\sim \Dist(0, \sigma_{w}^2/m), b \sim \Dist(0, \sigma_{b}^2) \nonumber
\end{align}
where the prior distribution $\Dist$ can be non-Gaussian. Because the
weight and bias parameters are subject to be i.i.d., and have zero
mean, the hidden units $h_i({x})$ are i.i.d. bounded random variables.
\comment{
\begin{align}
\Expect[w_i h_i(\bm{x})] &= \Expect[w_i] \Expect[h_i(\bm{x})] = 0 \label{eq:nngp:hidden:mean} \\
\Variance[w_i h_i(\bm{x})] &= \Expect[w_i^2] \Expect[ h_i(\bm{x})^2] = \sigma_{w}^2  \Expect[ h_i(\bm{x})^2] \label{eq:nngp:hidden:var} 
\end{align}
}
Following the Central Limit Theorem that, for $m$ i.i.d. random
variables with bounded mean and variance, the summation of them is a
Gaussian distribution when $m \rightarrow \infty$. Thus, we have
$f({x})$ as an approximate Gaussian distribution when the width of
output layer $m$ is large, as given in Eq.~(\ref{eq:nngp:dist}).
\begin{align}
    \label{eq:nngp:dist}
    f({x}) \sim \Norm(\Zero, \sigma_{b}^2 + \sigma_{w}^2   \Expect[ h_i({x})^2]) 
\end{align}
Likewise,
following the multi-dimensional Central Limit Theorem, any finite
collection $f({X}) = \{f({x}_1),\cdots, f({x}_N) \}$ have a
joint multivariate Gaussian distribution, which is exactly a GP.
\begin{align}
    f({X}) & \sim \Norm(\Zero, \Cov) \\
    \Cov & = \sigma_{w}^2 \Expect[\bm{\Phi} \bm{\Phi}^T] + \sigma_{b}^2 \Ident \label{eq:nngp:kernel}
\end{align}
This GP is the Neural Network Gaussian Process (NNGP). This reveals
priors over infinite wide neural network leads to an equivalence
to GP.  In the kernel function $\Cov$ of Eq.~(\ref{eq:nngp:kernel}),
$\bm{\Phi}$ is the design matrix given the parametric basis function
$\{h_1({x}) \cdots, h_m({x})\}$.  The difference between the
NNGP kernel (Eq.~(\ref{eq:nngp:kernel})) and the standard liner
model's kernel (Eq.~(\ref{eq:gp:joint:dist})) is that the NNGP kernel
needs to compute the expectation of the product of the design matrix
w.r.t. the prior distribution of the parameters, which is used to
define the basis function. This enables NNGP to be not only a model
ensemble in the space of the linear parameters ${w}$ but also in
the space of the basis function parameters ${w}^0$. By recursively
applying Central Limit Theorem, the kernel of deep neural network is
induced~\cite{DBLP:conf/iclr/LeeBNSPS18} in
Eq.~(\ref{eq:nngp:kernel:recusive}), where $f^{l}_j$ is the
pre-activation of the $j$-th hidden unit in the $l$-th layer and
$\bm{\Phi}^{l}$ is the design matrix defined by the $0$-th to $l$-th
layers of the neural network. The base case kernel $\Cov^0$ in
Eq.~(\ref{eq:nngp:kernel:base}) is equivalent to the kernel of BLR. 
\begin{align}
    \Cov^l &= \Expect[f^l_j({X}) f^l_j({X})] \nonumber \\ 
        &=  \sigma_{w}^2 \Expect_{f^{l - 1}_j \sim \Norm(\Zero, \Cov^{l - 1})}[\bm{\Phi}^{l - 1} (\bm{\Phi}^{l - 1})^{T}] + \sigma_{b}^2 \Ident \label{eq:nngp:kernel:recusive}\\
    \Cov^{0} &= \sigma_{w}^2{X}{X}^{T} + \sigma_{b}^2 \Ident \label{eq:nngp:kernel:base}
\end{align}
The NNGP kernel can be computed analytically under certain activation
functions $\act(\cdot)$, (e.g., the rectified linear function
$\relu$~\cite{DBLP:conf/nips/ChoS09}, the error function
$\errFunc$~\cite{DBLP:journals/neco/Williams98}).  To conduct
inference, as the model is a standard GP, we can get the
exact solution by Eq.~(\ref{eq:gp:pred}).


We discuss the complexity of NNGP. As a standard GP, exact 
prediction needs to compute the inverse of the kernel matrix in
$O(N^3)$, where $N$ is the number of training data points. Note that
the kernel matrix is in closed-form and the inversion only needs to
compute once in advance. For a new test data point, inference takes
vector-matrix multiplication in $O(N^2)$.  Quadratic time complexity
w.r.t. training data is an obstacle to deploying GP model.  But,
with the acceleration of parallel and GPU architecture, the training
and inference time is reasonable, as confirmed in our extensive
experimental studies. Note that GP is usually data-efficient.



\stitle{Model design for cardinality estimation}: To test query
set $Q$, we use the mean-squared-error (MSE) as the empirical loss
function of NNGP in Eq.~(\ref{eq:gp:exploss}), which is shown in
Eq.~(\ref{eq:loss:mse:mean})-(\ref{eq:loss:mse:square}).
\begin{align}
\mathcal{L}_{\kw{emp}}(Q) &= \frac{1}{|Q|} \sum_{q \in Q} \mathcal{L}_{\kw{emp}}(c(q), \widehat{c}(q)) \label{eq:loss:mse:mean} \\
\mathcal{L}_{\kw{emp}}(c(q), \widehat{c}(q)) &= |\log c(q) - \log
 \widehat{c}(q) |^2  =  \log^2 \frac{c(q)}{\widehat{c}(q)}  
\label{eq:loss:mse:square} 
\end{align}
To achieve an average low relative error, the target $c(q)$ and
prediction $\widehat{c}_(q)$ are transformed to logarithmic scale.
%
%
It is worth noting that minimizing the average of
$\log \frac{c(q)}{\widehat{c}(q)}$ (Eq.~(\ref{eq:loss:mse:mean})) is equivalent to minimize the
geometric mean of \Qerror, and minimizing the squared-error
(Eq.~(\ref{eq:loss:mse:square})) further imposes higher weights on larger
\Qerror over the average due to the
square~\cite{DBLP:journals/pvldb/DuttWNKNC19}.
When the empirical loss is squared loss, the prediction that minimizes the expected loss of Eq.~(\ref{eq:gp:exploss}) is the mean of the predictive distribution, i.e., Eq.~(\ref{eq:gp:mean}).

\section{Calibration of Uncertainty}
\label{sec:uncertainty}

In this section, we investigate the uncertainty calibration for
cardinality estimation by 
%
%
 NNGP in comparison with two existing DL approaches for uncertainty
 calibration that are applicable for regression tasks, namely,
 \DeepEnsemble and Bayesian Neural Network. We introduce them below.

\etitle{\DeepEnsemble} is a uniformly-weighted mixture model where
each mixture is a deep neural
network~\cite{DBLP:conf/nips/Lakshminarayanan17}.  Each neural network
treats one data point as a sample from a Gaussian distribution, in
predicting the mean and variance via the final layer of the neural
network.  Training one neural network $w$ is to minimize its
negative Gaussian log-likelihood in Eq.~(\ref{eq:gaussian:nll}), where
$\mu_{w}({x})$ and $\sigma_{w}^2({x})$ are its
predictive mean and variance, respectively.
\begin{align}
\label{eq:gaussian:nll}
- \log p_{w}(y | {x}) = \frac{\log \sigma_{w}^2({x})}{2} + \frac{(y - \mu_{w}({x}))^2}{2 \sigma_{w}^2({x})}
\end{align}
The ensemble prediction is approximated as a Gaussian distribution whose mean and variance are respectively the mean and variance of $M$ neural networks. 
\comment{
\begin{align}
\label{eq:deepensemble}
\mu_{*}(\bm{x}) &= \frac{1}{M} \sum_{m}  \mu_{\theta_m}(\bm{x}) \\
\sigma_{*}(\bm{x}) &= \frac{1}{M} \sum_{m} (\sigma_{\theta_m}^2(\bm{x}) + \mu_{\theta_m}^2(\bm{x})) - \mu_{*}^2(\bm{x}) 
\end{align}
}
The idea of \DeepEnsemble is simple whereas it needs to maintain $M$
neural networks explicitly. 

\etitle{Bayesian Neural Network (BNN)}, as the BDL model, quantifies
 the uncertainty for neural network by defining a prior belief $p({w})$
on its parameterization.  The prediction is a distribution
marginalizing over the posterior distribution $p({w} | Y,
{X})$ as shown in Eq.~(\ref{eq:ensemble:bayes}).  The computation
of this marginalization is approximated by variational inference where
$q({w})$ is a tractable variational distribution.  Gal et
al. in~\cite{DBLP:conf/icml/GalG16, DBLP:journals/corr/GalG15a}
propose an efficient inference that relates a Bernoulli variational
distribution to BNN via dropout training of the neural network.
In~\cite{DBLP:conf/icml/GalG16, DBLP:journals/corr/GalG15a}, inference
is done by training with a dropout before weight layers and by
performing dropout at test time, and the output distribution is
approximated by $T$ Monte Carlo forward passes with stochastic
parameter ${w}_t$ (Eq.~(\ref{eq:bnn:mcdropout})).
\comment{
\begin{align}
\label{eq:bnn:ensemble:bayes}
   p(\bm{y}^{*}| \bm{X}^{*}, \bm{y}, \bm{X}) &= \int p(\bm{y}^{*}|\bm{X}^{*}, \bm{w}) p(\bm{w}| \bm{y}, \bm{X}) d\bm{w} \\ \label{eq:bnn:vi}
& \approx \int p(\bm{y}^{*}|\bm{X}^{*}, \bm{w}) q(\bm{w}) d\bm{w} \\ 
& \approx \frac{1}{T} \sum_{t = 1}^{T} p(\bm{y}^{*}|\bm{X}^{*}, \bm{w}_t)
\label{eq:bnn:mcdropout}
\end{align}
}
\begin{align}
   p({y}^{*}| {x}^{*}, Y, {X}) 
 \approx \int p({y}^{*}|{x}^{*}, {w}) q({w}) d{w} 
 \approx \frac{1}{T} \sum_{t = 1}^{T} p({y}^{*}|{x}^{*}, {w}_t)
\label{eq:bnn:mcdropout}
\end{align}
%

\begin{figure}[t]
\centering
%
\begin{tabular}[t]{c}
\subfigure[\forest~\NNGP]{
\vspace*{-0.2cm}
\label{fig:cov:nngp:forest}
 \includegraphics[width=0.32\columnwidth]{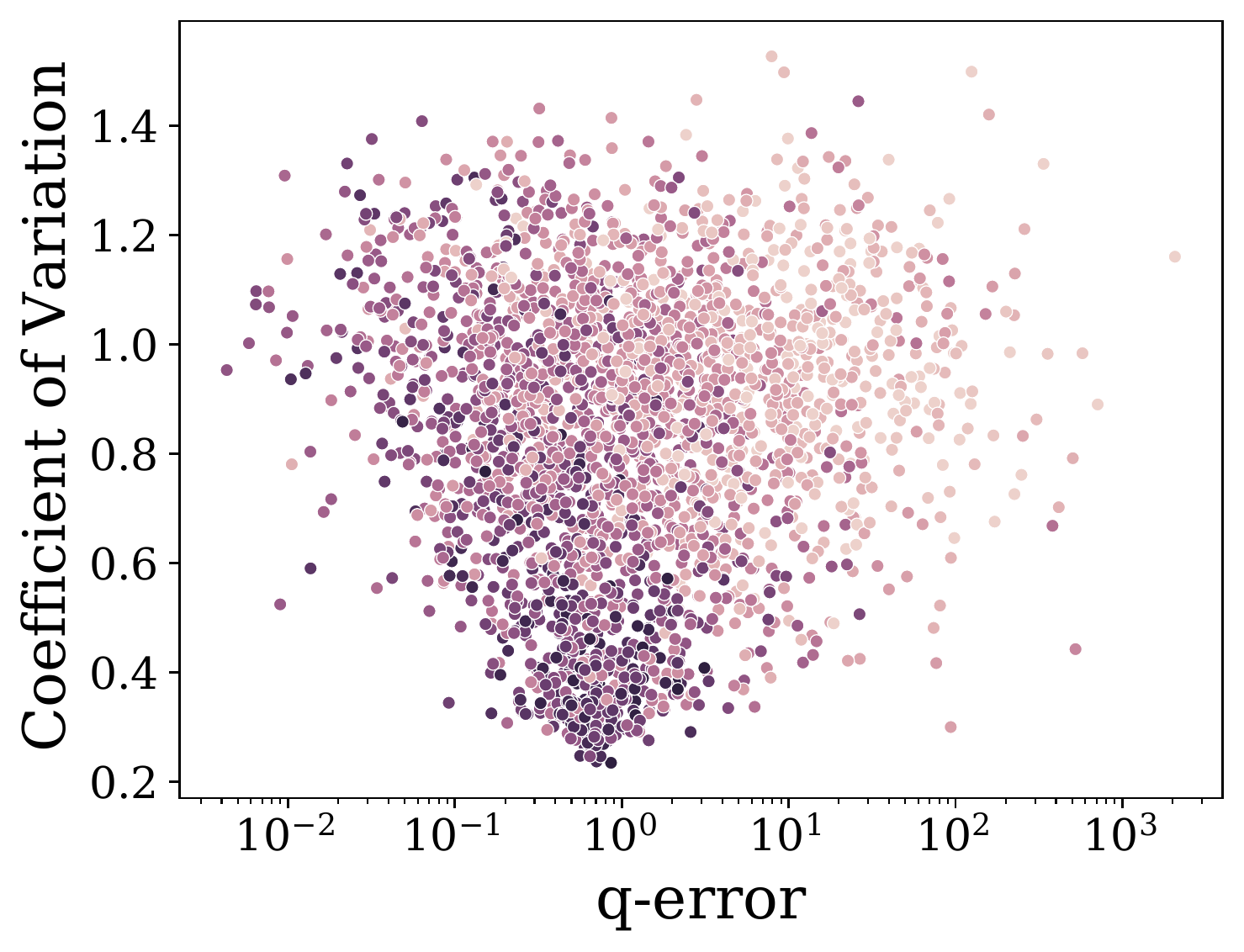}
\vspace*{-0.2cm}
}
\hspace*{-0.3cm}
\subfigure[\forest~\DeepEns]{
\label{fig:cov:deepens:forest}
\vspace*{-0.2cm}
 \includegraphics[width=0.32\columnwidth]{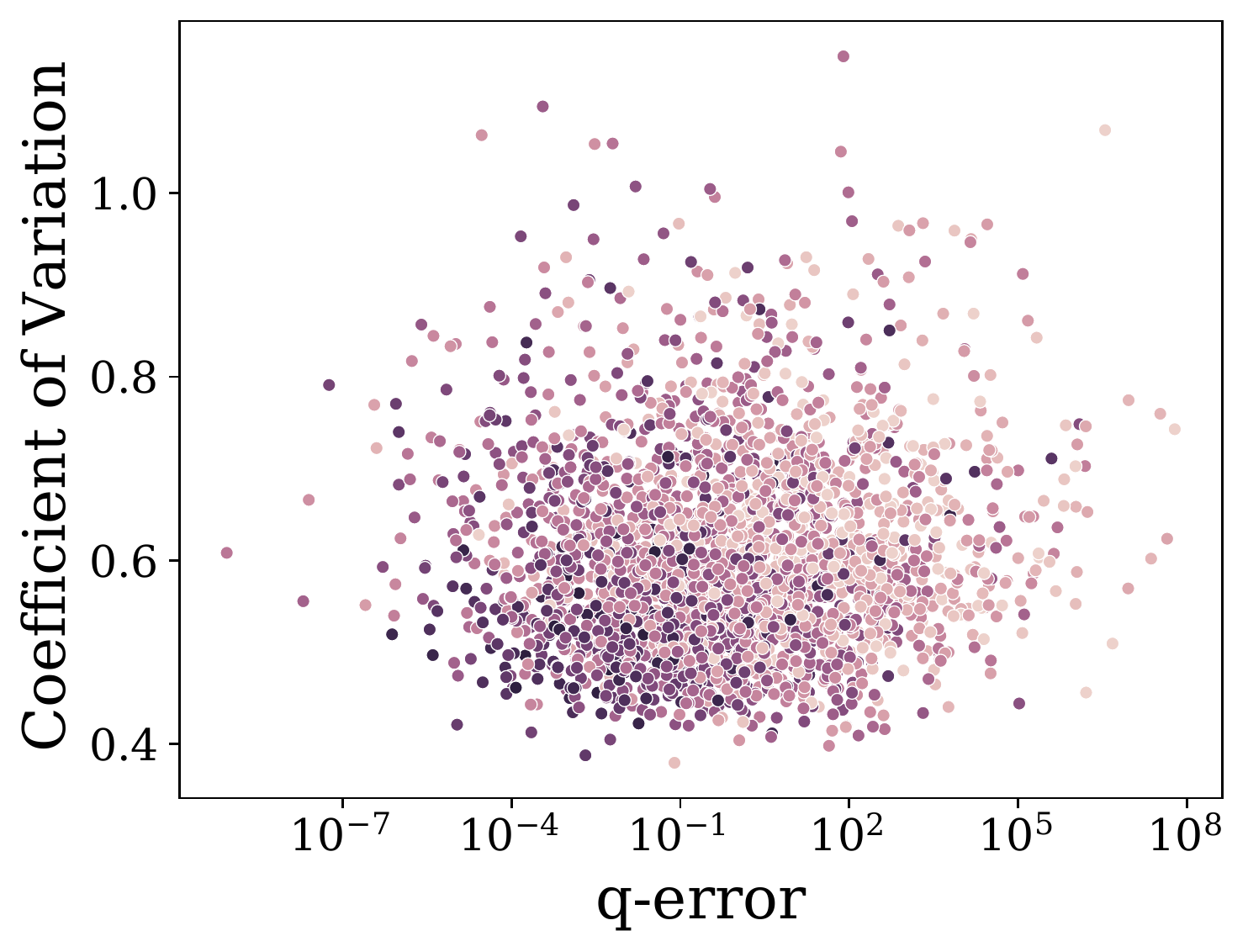}
\vspace*{-0.2cm}
}
\hspace*{-0.3cm}
\subfigure[\forest~\BNNMCD]{
\label{fig:cov:mcd:forest}
\vspace*{-0.2cm}
 \includegraphics[width=0.32\columnwidth]{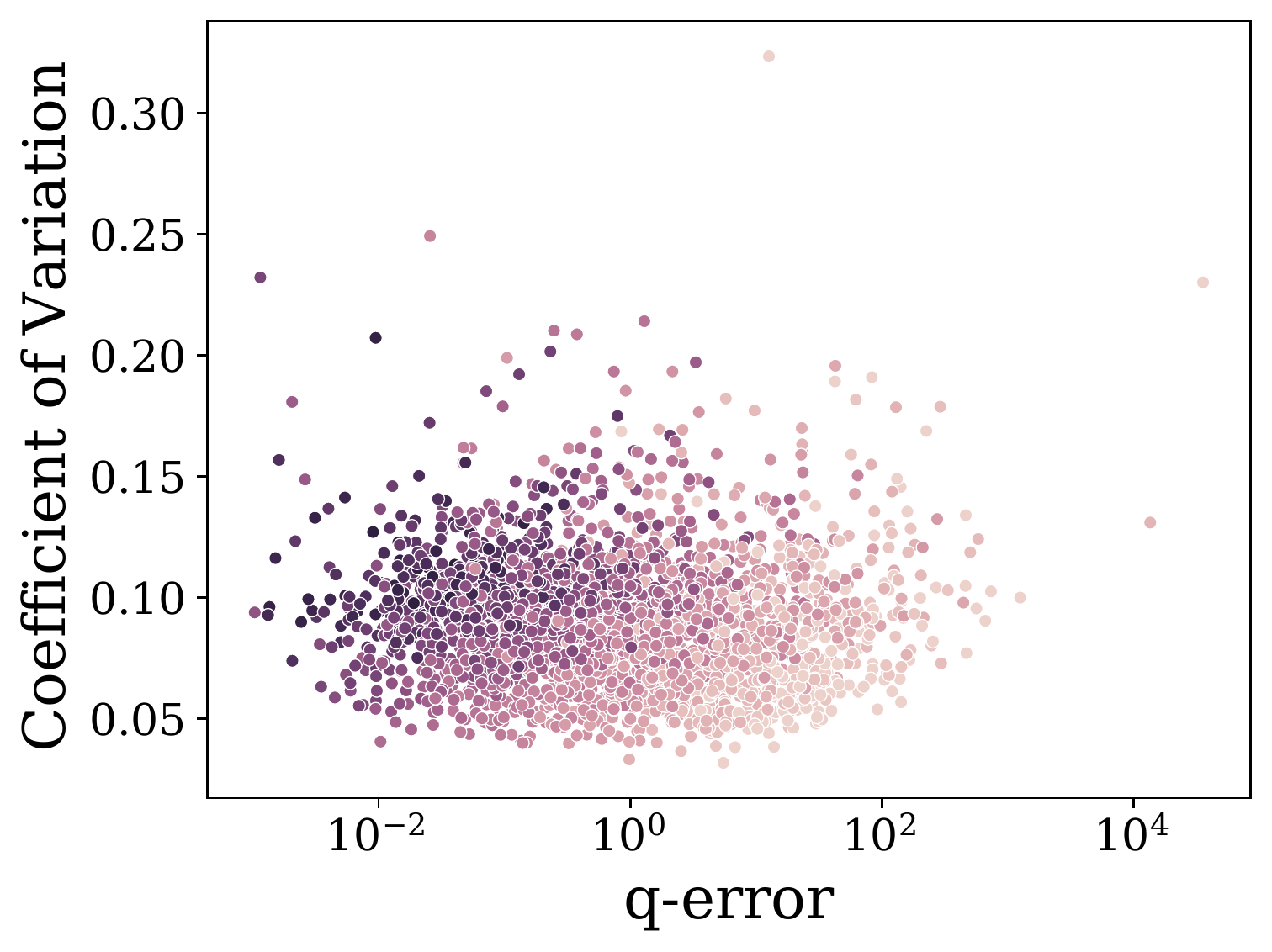}
 \vspace*{-0.2cm}
}
\end{tabular}

\begin{tabular}[t]{c}
\subfigure[\tpcds~\NNGP]{
\vspace*{-0.2cm}
\label{fig:cov:nngp:tpcds}
 \includegraphics[width=0.32\columnwidth]{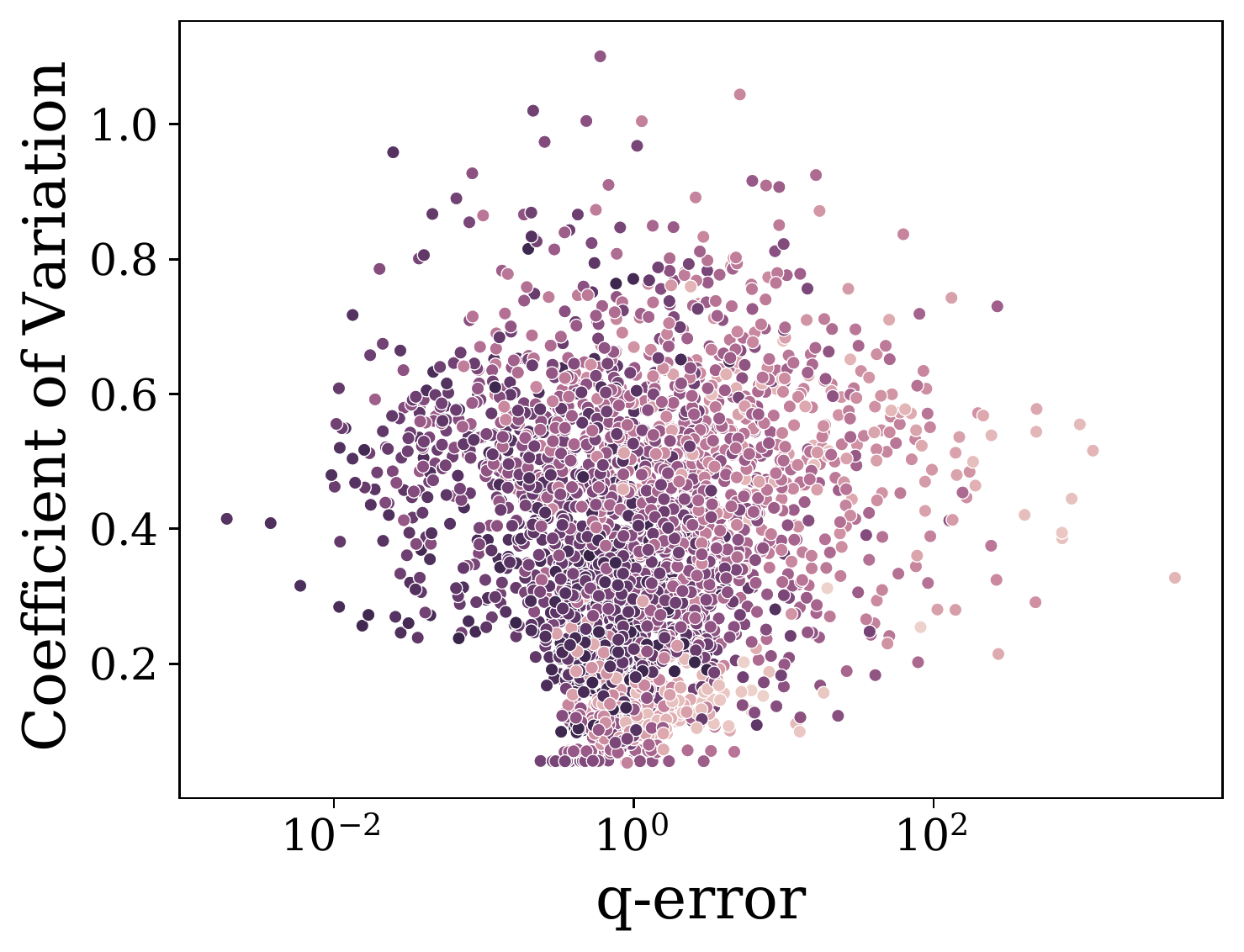}
\vspace*{-0.2cm}
}
\hspace*{-0.3cm}
\subfigure[\tpcds~\DeepEns]{
\label{fig:cov:deepens:tpcds}
\vspace*{-0.2cm}
 \includegraphics[width=0.32\columnwidth]{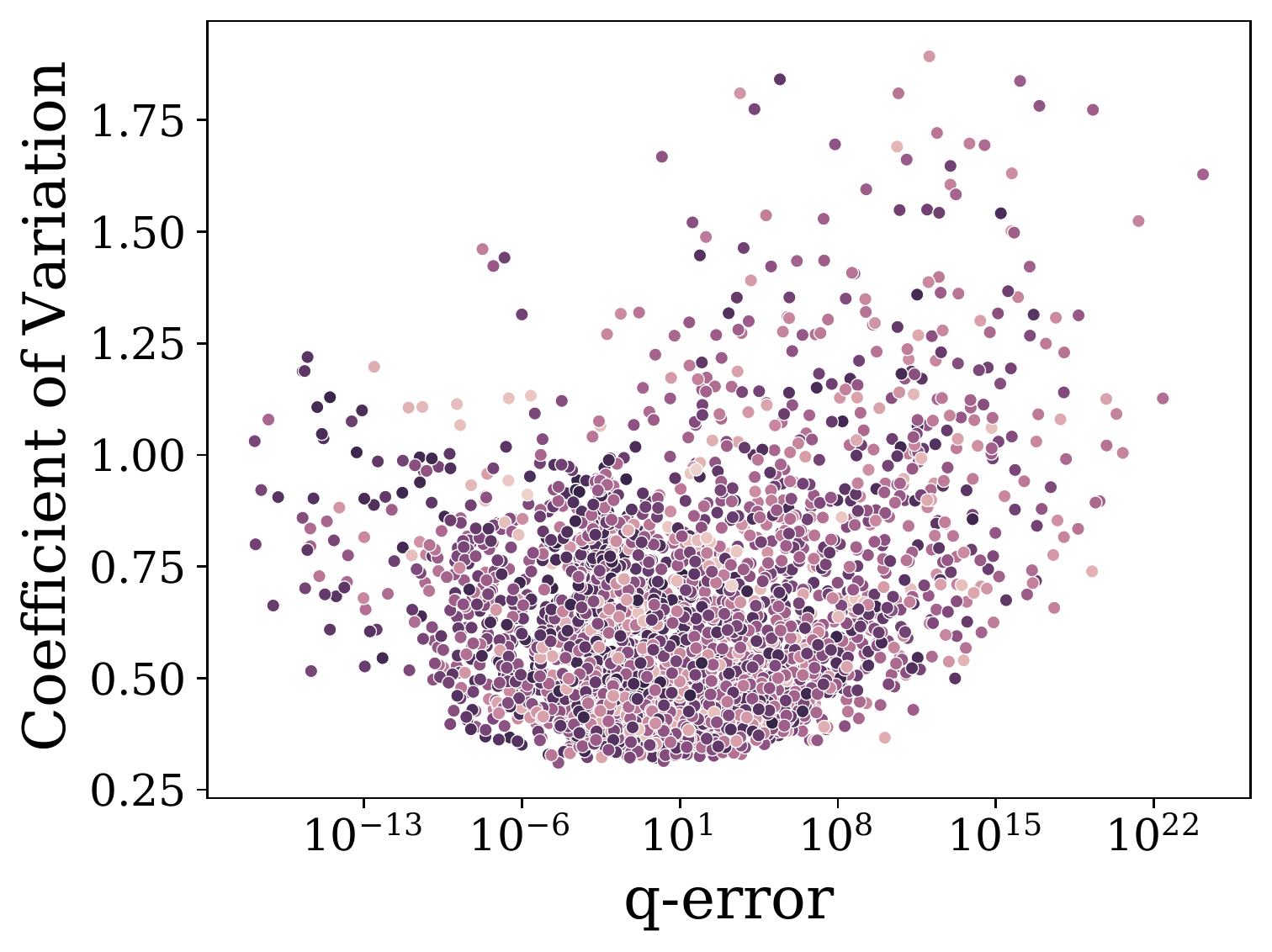}
\vspace*{-0.2cm}
}
\hspace*{-0.3cm}
\subfigure[\tpcds~\BNNMCD]{
\label{fig:cov:mcd:tpcds}
\vspace*{-0.2cm}
 \includegraphics[width=0.32\columnwidth]{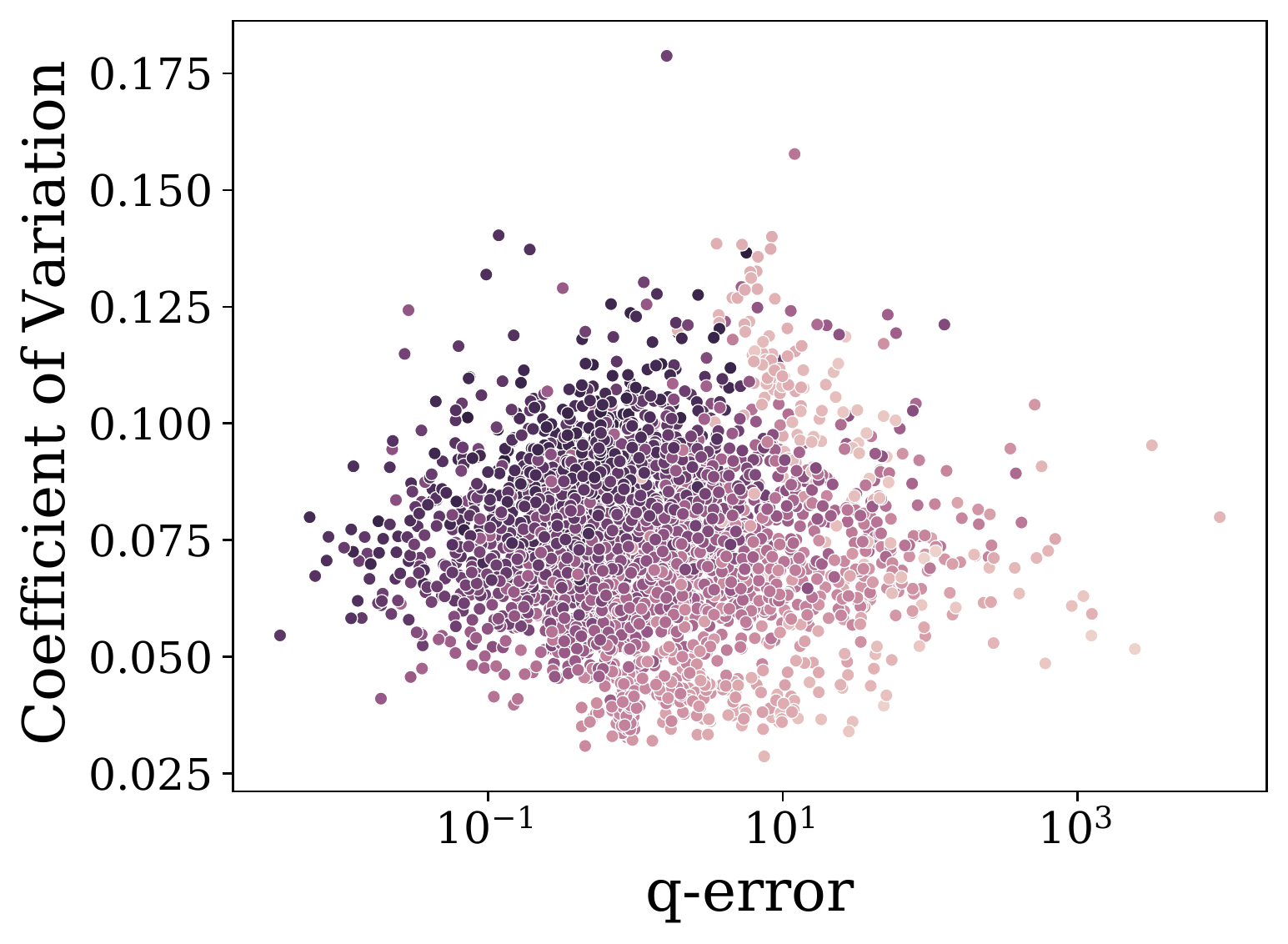}
\vspace*{-0.2cm}
}
\end{tabular}
\vspace*{-0.3cm}
\caption{Visualization of Estimation Uncertainty}
\label{fig:cov:visual}
\end{figure}

We have implemented above two techniques for the lightweight neural
network estimator~\cite{DBLP:journals/pvldb/DuttWNKNC19}, as \DeepEns
and \BNNMCD respectively. The estimator is a two-layer multilayer perceptron with 512 hidden units.
The \DeepEns ensembles 5 estimators  
and \BNNMCD uses a dropout probability of 0.5 and 2,000 forward passes
for the prediction.

Fig.~\ref{fig:cov:visual} visualizes the uncertainty and the \Qerror
for thousands of queries on \forest and \tpcds, where the points
denote testing queries.  The uncertainty is represented by coefficient
of variation, which is the standard deviation normalized by the mean, to
quantify the dispersion of the predictive distribution.  Ideally, the
uncertainty should be highly correlated with the prediction
error. That is, the larger the \Qerror, the larger the uncertainty.
\NNGP estimator shows such behavior.  For the queries with larger
\Qerror, their coefficient of variation of \NNGP estimator is large in
Fig.~\ref{fig:cov:nngp:forest} and Fig.~\ref{fig:cov:nngp:tpcds},
where the scatter plots are in a V-shape.
For \DeepEns and \BNNMCD, this V-shape distribution is not obvious.
The flat bottom in the scatter plots indicates the estimators are
still over-confident on some queries whose \Qerror is far away from
1.0.  Furthermore, we also observe that \DeepEns still has a relative
worse performance w.r.t. prediction accuracy, compared with \NNGP and
\BNNMCD.  And \BNNMCD faces the risk of unstable training as the
variational inference aims to optimize a lower bound of the original
training loss.  All the three estimators make some conservative
predictions in the sense that the actual \Qerror is small whereas the
uncertainty is large. It reflects the fact that all of them are
possible to process queries out-of-distribution, via different
smoothing techniques on their prediction.

In terms of efficiency, \DeepEns needs to train and maintain multiple
neural networks explicitly, and \BNNMCD needs to conduct a large
number of forward passes in the testing phase.  It is impractical to
deploy these two techniques on high-end DL estimators, (e.g., the
Tree LSTM estimator~\cite{DBLP:journals/pvldb/SunL19}) in a DBMS.  In \cref{sec:exp}, we show that the 
training and prediction of \NNGP are faster than wide neural network,
thus is faster than the \DeepEns and \BNNMCD counterparts.

\section{Experimental Studies}
\label{sec:exp}
In this section, we give
the test setting (\cref{sec:exp:setup}), and report the
extensive experiments in the following facets:
\ding{172} Compare the accuracy of NNGP estimator with state-of-the-art ML/DL estimators (\cref{sec:exp:acc}).
\ding{173} Test the training and prediction efficiency of NNGP estimator and verify it is a lightweight DL-based estimator (\cref{sec:exp:time}).
\ding{175} Validate the robustness of NNGP estimation regarding small number of training queries and unbalanced workloads (\cref{sec:exp:robust}) 
\ding{174} Study the application of the uncertainty that NNGP provides in active learning, which all the DL-based estimators do not support (\cref{sec:exp:active}).

\subsection{Experimental Setup}
\label{sec:exp:setup}
\stitle{Baseline Approaches.}
To comprehensively evaluate the effectiveness and efficiency of NNGP estimator, we compare it with 8  estimators including 5 query-driven learned estimators, 2 data-driven learned estimators and 1 traditional estimator as follows.
\ding{202} Neural Network (\NN)~\cite{DBLP:journals/pvldb/DuttWNKNC19} is the standard fully-connected neural network with $\relu$ activation. 
\ding{203} Gradient Boosting Decision Tree (\GBDT)~\cite{DBLP:journals/pvldb/DuttWNKNC19, dutt2020efficiently} is the  ensembling decision regression trees by gradient boosting. 
\ding{204} Multi-set Convolutional Neural Network (\MSCN)~\cite{DBLP:conf/cidr/KipfKRLBK19} firstly embeds the table set, join set and predicate set by 3 separate multilayer perceptrons as the set convolutions, respectively, and then concatenate the 3 embeddings to a long vector and feed it into a final output network. 
\ding{205} Tree LSTM (\TLSTM)~\cite{DBLP:journals/pvldb/SunL19} is a high-end DL model originally designed for cost and cardinality estimation, by taking tree-structured query plans as its input. 
It is composed of three stacked layers, the embedding layer, representation layer and estimation layer. The embedding layer embeds operations (join and table scan), predicates, metadata of leaf nodes of the plan tree into vector representations. The embedding layer aggregates the representation of each node on the tree in a bottom-up fashion by Tree-Structured LSTM~\cite{DBLP:conf/acl/TaiSM15}. 
The estimation layer is a fully-connected neural network with $\sigmoid$ activations that finally outputs the predictions.
A detailed formulation of the model can be found in \cite{DBLP:journals/pvldb/SunL19}. 
\ding{206} \DeepDB~\cite{DBLP:journals/pvldb/HilprechtSKMKB20} (\DeepDBExp) is an unsupervised learning, data-driven  estimator that uses Relational Sum-Product Network (RSPN)~\cite{DBLP:conf/aaai/NathD15} to model the joint distribution of a relation. To support join queries, an ensemble of RSPNs or a joint RSPN is built, and the choice is based on independent test on the relations. SQL queries are complied into probabilistic queries on the RSPN. 
\ding{207} \Naru~\cite{DBLP:journals/pvldb/YangLKWDCAHKS19}/\NeuroCard~\cite{DBLP:journals/corr/abs-2006-08109} (\NeuroCardExp) factorizes the joint distribution into conditional distributions and adopts deep autoregressive models such as MADE~\cite{DBLP:conf/icml/GermainGML15} or Transformer~\cite{DBLP:conf/nips/VaswaniSPUJGKP17} to approximate the joint distribution. \NeuroCard further extends \Naru to support full outer joins.
To predict the cardinalities, progressive sampling is conducted over the density model. 
\ding{208} {\sl PostgreSQL} estimator (\Postgres) is a build-in statistical estimator.  Estimated cardinality is obtained from the \textsf{EXPLAIN} command of {\sl PostgreSQL}.
\ding{209} Gaussian Process with radial basis kernel function (\GPRBF) is compared as a typical GP baseline.
All the above learned estimators only support PK/FK joins.  

\stitle{Implementation and Settings.}
The NNGP estimator is built on \NTK~\cite{neuraltangents2020}, which is based on Google~\JAX~
\cite{jax2018github}. The empirical zero-mean Gaussian prior is imposed on the weights of neural networks and $\relu$ is used as the nonlinear activation, leading to a closed-form kernel function for Bayesian inference. 
There is no extra cost paid for \NNGP hyper-parameters tuning.
\GBDT estimator is implemented by \XGBoost~ \cite{caDBLP:conf/kdd/ChenG16} and a tree ensemble contains 32 trees. 
\GPRBF is implemented by the exact GP regressor of {\sl scikit-learn}.
For \NeuroCardExp, we follow all the hyper-parameter configurations in its implementation based on MADE as the deep autoregressive model. 
\NN, \MSCN, and \TLSTM estimators are implemented by {\sl PyTorch} 1.6 ~\cite{pytorch}.
We use the Adam optimizer with a decaying learning
rate to train these models. For different datasets, the main hyper-parameters
for training are tuned in their empirical range: learning rate $\in [10^{-3}, 10^{-4}]$, epochs $\in \{50, 80, 100\}$, mini-batch size $\in \{16, 32, 64\}$, L2 penalty of Adam $\in [10^{-3}, 10^{-5}]$.
For \NN, \GBDT, \GPRBF and \NNGP, their encodings of input are same, as the introduction in \cref{sec:ps}. Other models have their own input encoding due to their specific model design. 
Particularly, as \TLSTM takes a tree-structured plan as input, we generate a left-deep tree for each join query following a total order of the relations. 
Since we only perform one task of cardinality estimation, one-hot physical operator encoding is simplified to the corresponding one-hot logical operator encoding. 
Both training and prediction are performed on a Linux server with
32 Intel(R) Xeon(R) Silver 4215 CPUs and 128G RAM.

\comment{
\begin{table}[t]
{ \small 
\caption{The Hyper-parameter Configuration}
\label{tbl:hyperpara}
\begin{center}
    \begin{tabular}{|l|c|} \hline
    {\bf Hyper-parameters}  & {\bf Values} 
     \\\hline\hline
  	 learning rate & $10^{-3} \sim 10^{-4}$\\ \hline
	 mini-batch size   &  ${16, 32, 64}$ \\ \hline 
     epochs & $\{50, 80, 100, 150\}$ \\\hline
     weight decay (L2 penalty) & $10^{-3} \sim 10^{-5}$ \\ \hline
     \# hidden units &  \{64, 128, 256\}  \\ \hline 
     \end{tabular}
\end{center}
}
\end{table}
}

\begin{table}[t]
{\footnotesize
\caption{Query Sets}
\label{tbl:querysets}
\vspace*{-0.4cm}
\begin{center}
    \begin{tabular}{|l r r r r|} \hline
     {\bf Type} & {\bf Dataset}   & {\bf \# Queries} & {\bf \# Select/Join Cond. } & { \bf Range of $c(q)$}
    \\ \hline\hline                                               
   Single Rel. & \forest   & 18,000 & \{2, $\cdots$, 10\} & $[10^0, 10^6]$  \\ \hline 
   Single Rel. & \higgs  & 12,000 & \{2, $\cdots$, 7\} &  $[10^{0}, 10^8]$   \\\hline \hline
   Join & \tpch & 16,000 & \{0, 1, 2, 3\} & $[10^{0}, 10^{7}]$  \\\hline
   Join & \tpcds  & 15,000 & \{0, 1, 2, 3, 4\} & $[10^{0}, 10^{7}]$  \\\hline
   \end{tabular}
\end{center}
}
\end{table}

\stitle{Datasets.} We use 4 relational datasets, 2 for single relation range queries, 2 for multi-relation join queries. 
\forest~\cite{Dua:2019} originally contains 54 attributes of forest cover type. Following \cite{DBLP:journals/pvldb/DuttWNKNC19, DBLP:journals/vldb/GunopulosKTD05}, we use the first 10 numerical  attributes. The relation has about 581K of rows. 
\higgs~\cite{Dua:2019} is a physical dataset contains 7 high-level kinematic attributes of particles, collected by detectors and post-processed by scientific functions. The relation has 11M rows.   
\tpch (1 GB). We use the relations \textsf{supplier}, \textsf{orders}, \textsf{part} and \textsf{lineitem}. There are 3 PK/FK join conditions. 
\tpcds (2 GB). We use the relations \textsf{store}, \textsf{item}, \textsf{customer} and \textsf{promotion}, \textsf{store-sales}, where \textsf{store-sales} is the factual relation and others are the dimensional relations. There are 5 PK/FK join conditions and the schema has a cycle. 

\stitle{Queries.}
We construct large query workloads in the following way.
For single relation \forest and \higgs,  we generate query sets with the number of select conditions varying from 2 to $D$ where $D$ is the number of attributes, and generate 2,000 queries for each subset. 
To generate a query of $d$ selection conditions, first we uniformly sample $d$ attributes from all the $D$ attributes, then uniformly sample each attribute by the data-centric distribution following ~\cite{DBLP:journals/pvldb/DuttWNKNC19}. 
For join queries, we test and report the existing baselines supported query type, i.e., PK/FK join without selection conditions on the join attributes. We generate query sets with the number of PK/FK joins, t, varying from 0 to $|T| - 1$, where $|T|$ is the number of involved relations.   
To generate a query of $t~(t > 0)$ joins, firstly a starting relation is uniformly sampled, then the query is constructed by traversing from the starting relation over the join graph in $t$ steps.
Here, for each relation of the sampled join query, additional selection conditions are drawn independently.  
For each $t$, 4,000 and 3,000 queries are generated for \tpch and \tpcds, respectively. 
We only preserve unique queries with nonzero cardinality. 
Table~\ref{tbl:querysets} summarizes the 4 corresponding query sets.

\subsection{Accuracy}
\label{sec:exp:acc}

\begin{figure}[t]
\centering
\begin{tabular}[t]{c}
\centering
 \includegraphics[width=0.8\columnwidth]{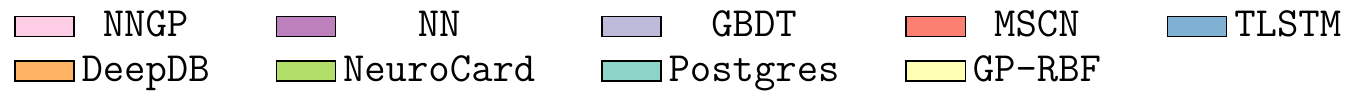}
 \vspace*{-0.3cm} 
\end{tabular}
\begin{tabular}[t]{c}
\vspace*{-0.2cm}
\subfigure[\forest]{
\label{fig:exp:acc:forest}
 \includegraphics[width=0.9\columnwidth]{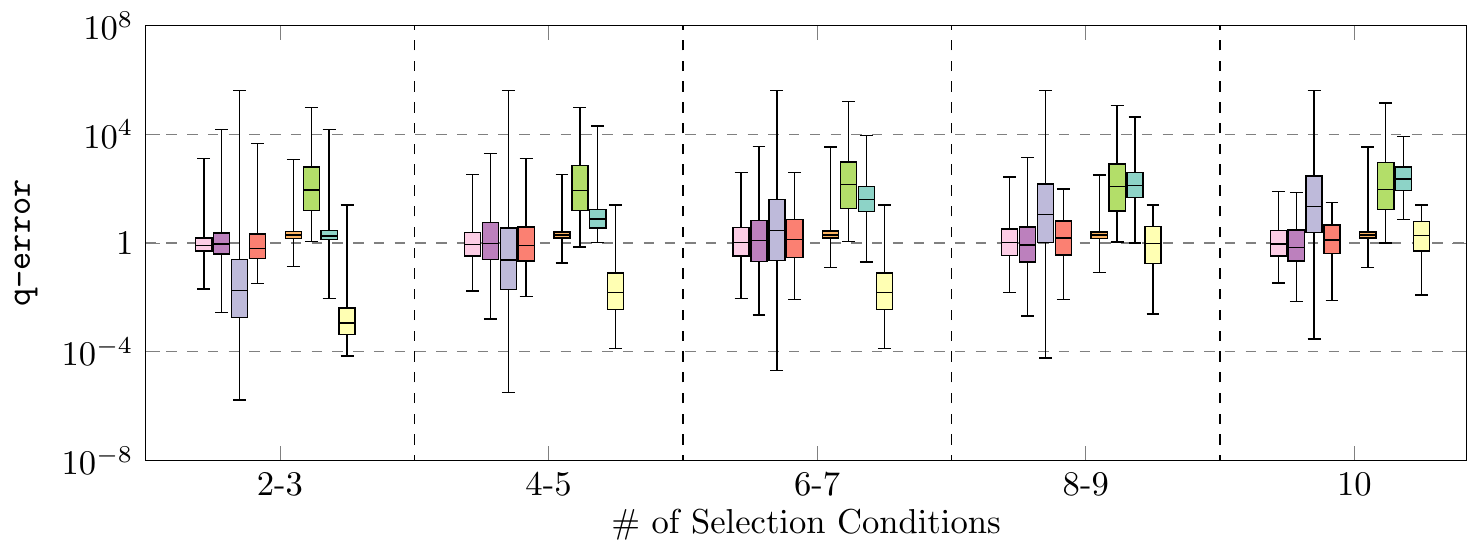}
}
\end{tabular}
\begin{tabular}[t]{c}
\vspace*{-0.2cm}
\subfigure[\higgs]{
\label{fig:exp:acc:higgs}
 \includegraphics[width=0.9\columnwidth]{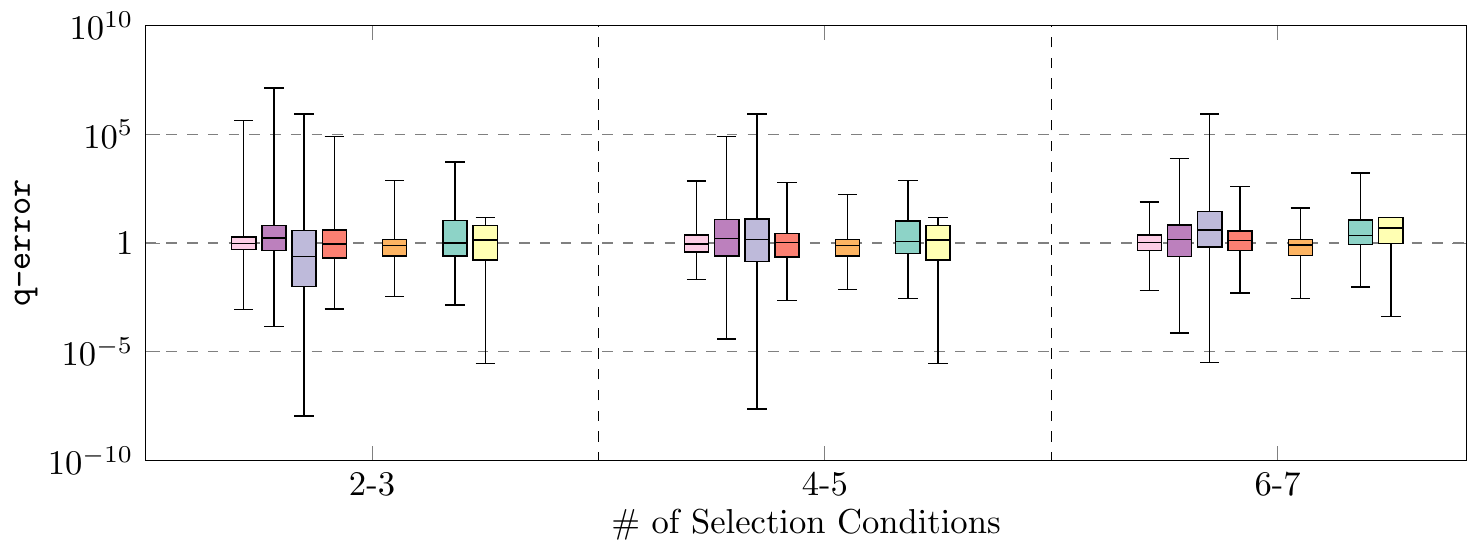}
}
\end{tabular}
\begin{tabular}[t]{c}
\vspace*{-0.2cm}
\subfigure[\tpch]{
\label{fig:exp:acc:tpch}
 \includegraphics[width=0.9\columnwidth]{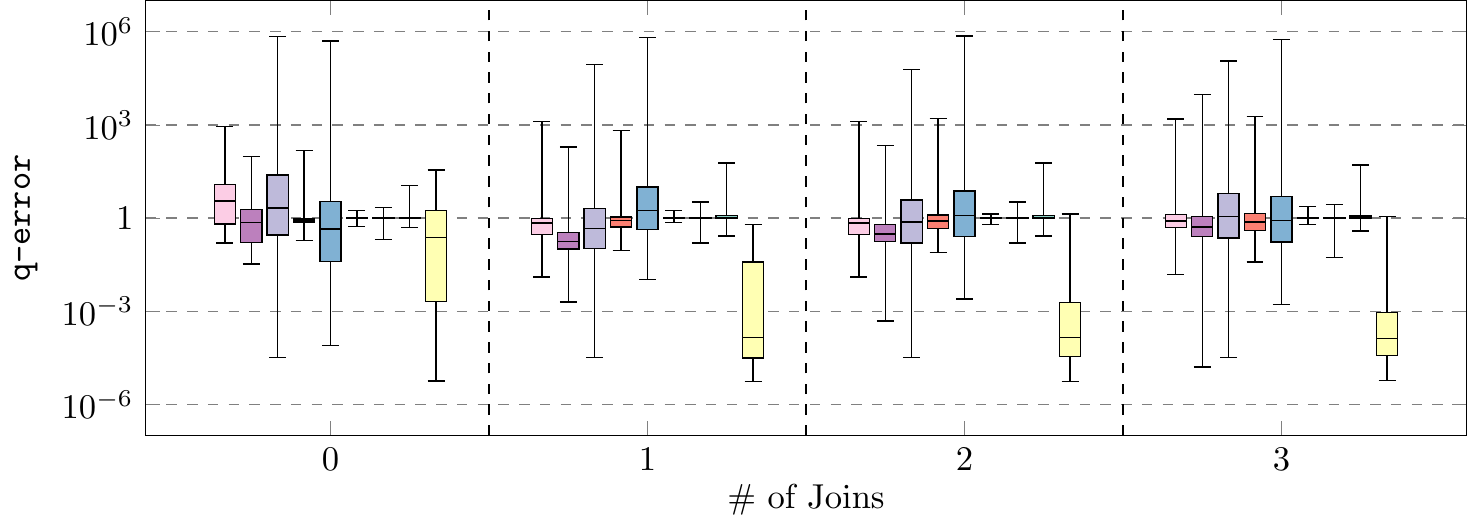}
}
\end{tabular}
\begin{tabular}[t]{c}
\vspace*{-0.2cm}
\subfigure[\tpcds]{
\label{fig:exp:acc:tpcds}
 \includegraphics[width=0.9\columnwidth]{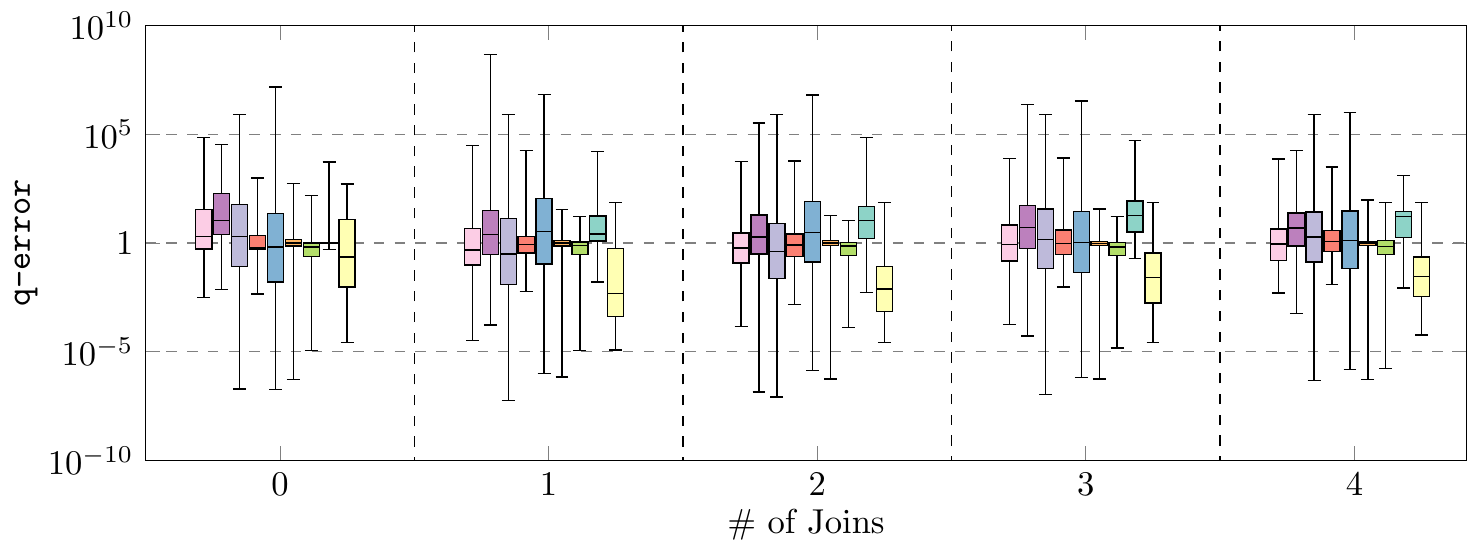}
}
\end{tabular}
\vspace*{-0.2cm}
\caption{Query Evaluation Accuracy}
\label{fig:exp:acc}
\end{figure}

We investigate the estimation accuracy of NNGP estimator. 
For the query-centric approaches, i.e., \NNGP, \NN, \GBDT, \MSCN, \TLSTM and \GPRBF, we split the queries into 60\% for training, 20\% for validation and 20\% for testing. 
The split is conducted by stratified sampling on the subsets with different numbers of selection/join conditions. 

Fig.~\ref{fig:exp:acc} presents the statistical distribution of \Qerror of \NNGP compared with the 8 baselines. 
In general, \NNGP, \MSCN and \DeepDBExp are the top three best performed estimators. There is no an overall best estimator that consistently outperforms others under all the test circumstances. 
Due to the infinite wide hidden layer, the performance of \NNGP consistently surpasses the finite wide \NN.
The performance advantage of \NNGP is mainly reflected in queries of single relation  (Fig.~\ref{fig:exp:acc:forest}, \ref{fig:exp:acc:higgs}), and the 25\%-75\% quantile of \Qerror is within 1.5. 
For the join queries  (Fig.~\ref{fig:exp:acc:tpch}, \ref{fig:exp:acc:tpcds}), we speculate the join encoding and the multilayer perceptrons based kernel of \NNGP are simple to model the complex join queries. 
In contrast, the performance advantage of \MSCN is reflected in the join queries rather than queries on single relation.
Due to satisfaction the behavior logics, \DeepDBExp achieves a promising 25\%-75\% quantile of \Qerror. However, we observe that there are many over/underestimated queries for \forest (Fig.~\ref{fig:exp:acc:forest}) and \tpcds (Fig.~\ref{fig:exp:acc:tpcds}), as the model would factorize intertwined attributes under the conditional independent assumption. 

We discuss other baseline approaches.  
Although \TLSTM is a high-end DL estimator with complex layers, its performance is not promising for a sole cardinality estimation task. 
We speculate \TLSTM is mainly applicable for real cost estimation tasks when more features about the physical operators and meta-data of the physical database are fed into the model. 
Its original implementation learns \TLSTM by multi-task learning for the physical cost and the cardinality. However, the tree-structured plan directly influences the physical cost but not the cardinality. 
The simple GP baseline, \GPRBF, underfits the training queries, and therefore it cannot be used as a cardinality estimator. It indicates the representation learning provided by DL is critical for a complicated learning task. 
\Postgres uses single column statistics and assumes the columns are independent. This conventional statistical approach performs well on the benchmark \tpch but leads to large estimation bias on other datasets, particularly for complex queries.

\subsection{Efficiency}
\label{sec:exp:time}
\begin{figure}
\centering
 \includegraphics[width=0.9\columnwidth]{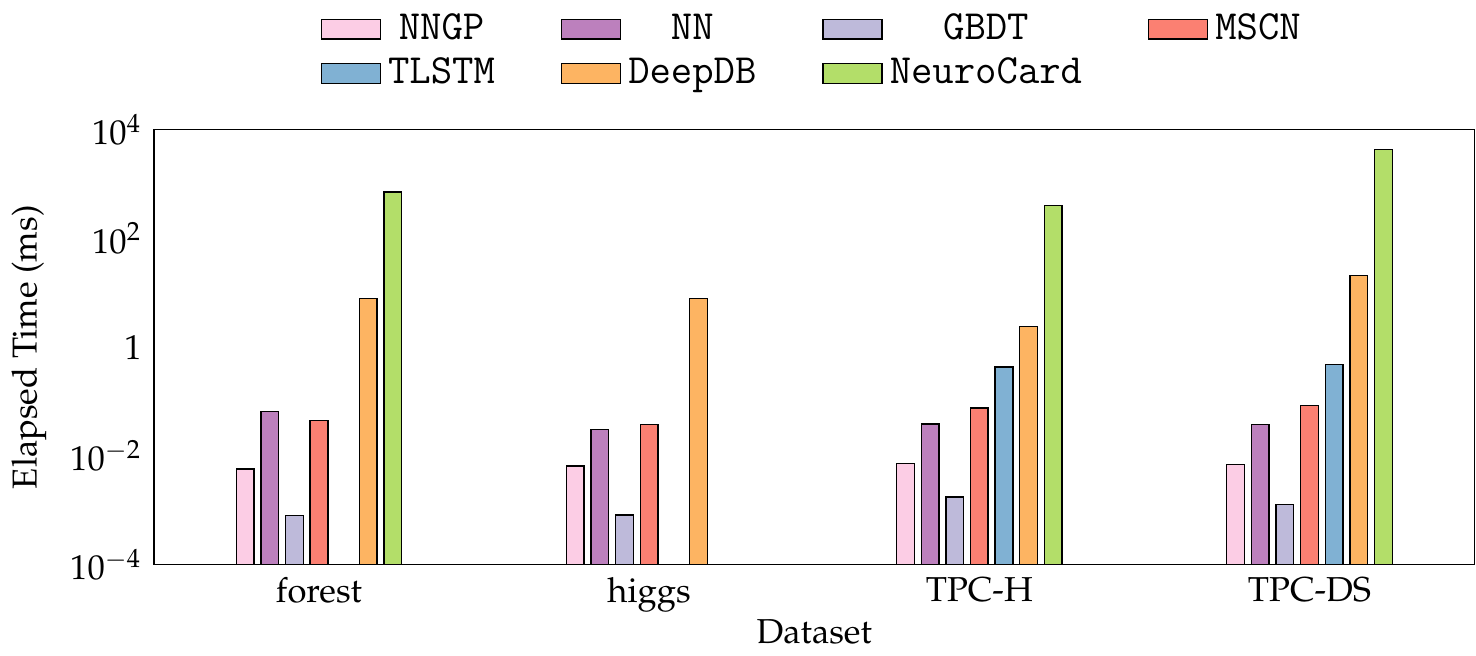}
\vspace*{-0.3cm}
 \caption{Query Evaluation Latency (ms)}
 \label{fig:exp:efficiency}
\end{figure}

We compare the prediction time of \NNGP estimators with the baseline approaches. Fig.~\ref{fig:exp:efficiency} presents the average elapsed time of the whole test query set on our 32-core CPU. 
In general, the lightweight \GBDT estimator achieves the fastest prediction and our \NNGP estimator is the second best. 
\XGBoost is a well-optimized learning system for \GBDT. 
The inference of \NNGP is a linear combination of the ground-truth cardinalities, where the weights are the closed-form kernels of Eq.~(\ref{eq:nngp:kernel:recusive}). 
The prediction of \DeepDBExp is 2-3 orders of magnitude slower than that of supervised learning approaches. 
\DeepDBExp needs to perform a bottom-up pass on the tree-structured deep sum-product network. 
Estimation on \NeuroCardExp is quite slow as one query needs thousands of sampling on the model to perform a Monte Carlo integration. We noticed that in its original paper~\cite{DBLP:journals/corr/abs-2006-08109}, the GPU executed prediction is up to 2 orders of magnitude faster than our CPU execution. 
That means GPU acceleration is necessary to employ \NeuroCardExp as an estimator available for DBMS. However,  \NeuroCardExp also consumes large GPU memory. Training on \forest is out of memory on a 16GB Nvidia V100.
In terms of easy use for DBMS, lightweight estimators like our \NNGP have a great advantage. 
For \NNGP, \NN and \GBDT, the prediction time for queries with different number of selection/join conditions are equal, as the input vectors are of equal length. 
For other estimators, the more complex the query, the longer the prediction time.
The prediction time of \DeepDBExp and \NeuroCardExp is also related to the database schema. In general, complex schema incurs a larger model and longer prediction time. 

\begin{figure}[t]
\centering
\begin{tabular}[t]{c}
 \includegraphics[width=0.4\columnwidth]{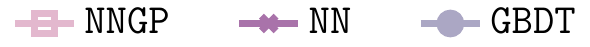}
\vspace*{-0.2cm}
\end{tabular} 
\begin{tabular}[t]{c}
\subfigure[\forest]{
\label{fig:exp:scale:time:forest}
 \includegraphics[width=0.4\columnwidth]{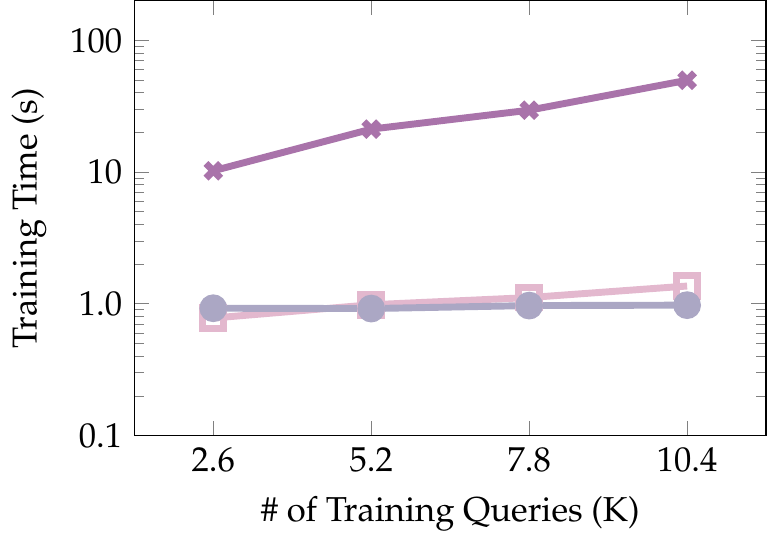}
}
\subfigure[\forest]{
\label{fig:exp:scale:memory:forest}
 \includegraphics[width=0.385\columnwidth]{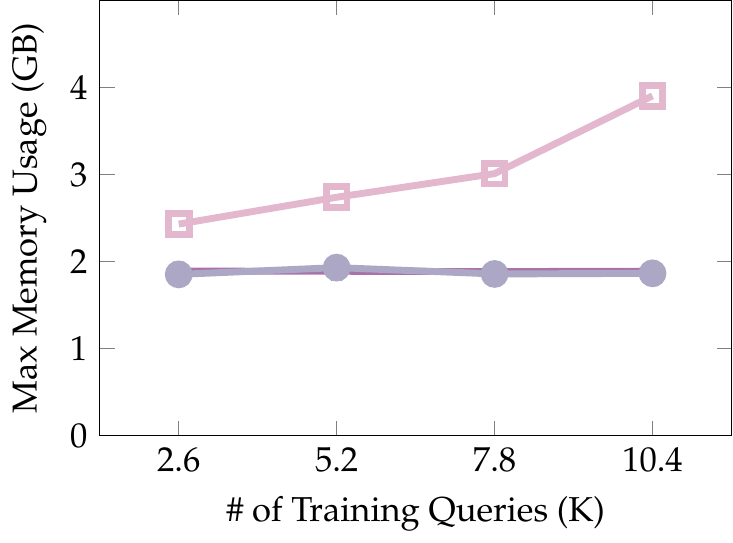}
}
\end{tabular}
\vspace*{-0.2cm} 
\begin{tabular}[t]{c}
\subfigure[\higgs]{
\label{fig:exp:scale:time:higgs}
 \includegraphics[width=0.4\columnwidth]{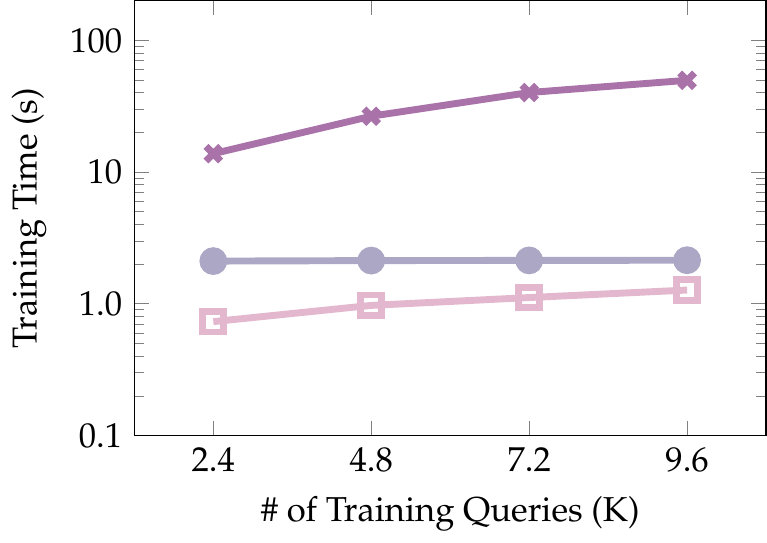}
}
\subfigure[\higgs]{
\label{fig:exp:scale:memory:higgs}
 \includegraphics[width=0.385\columnwidth]{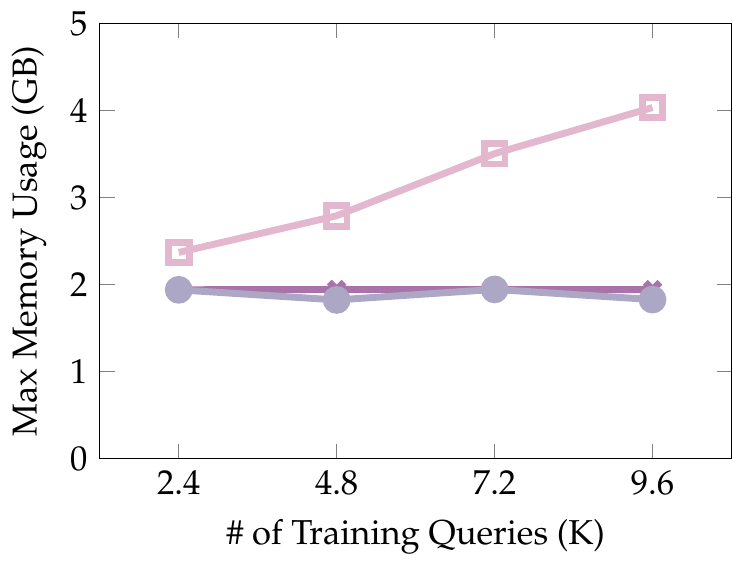}
}
\end{tabular}
\vspace*{-0.2cm} 
\begin{tabular}[t]{c}
\subfigure[\tpch]{
\label{fig:exp:scale:time:tpch}
 \includegraphics[width=0.4\columnwidth]{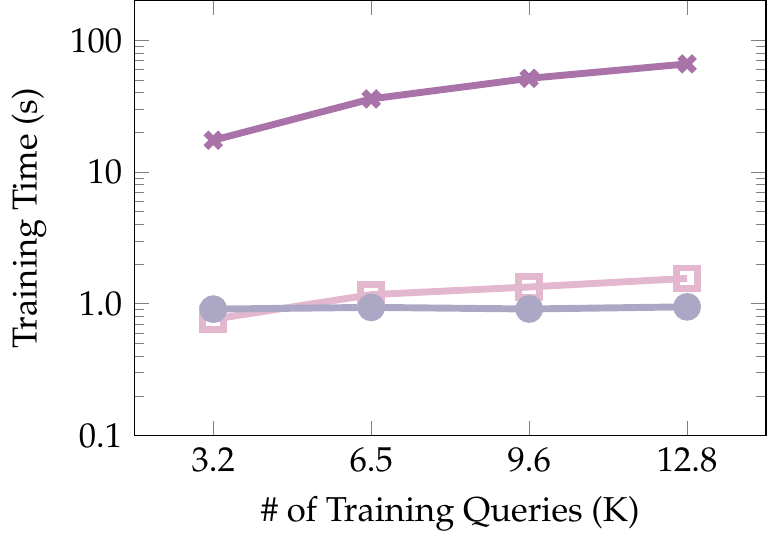}
}

\subfigure[\tpch]{
\label{fig:exp:scale:memory:tpch}
 \includegraphics[width=0.385\columnwidth]{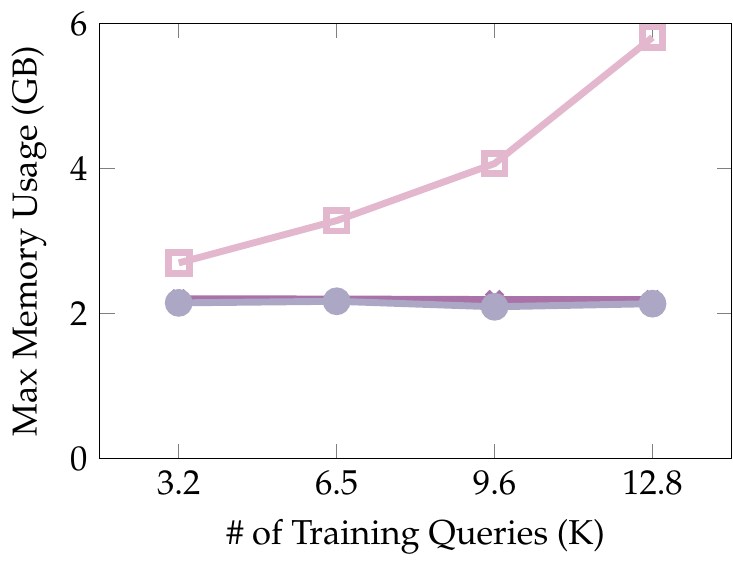}
}
\end{tabular}
\begin{tabular}[t]{c}
\subfigure[\tpcds]{
\label{fig:exp:scale:time:tpcds}
 \includegraphics[width=0.4\columnwidth]{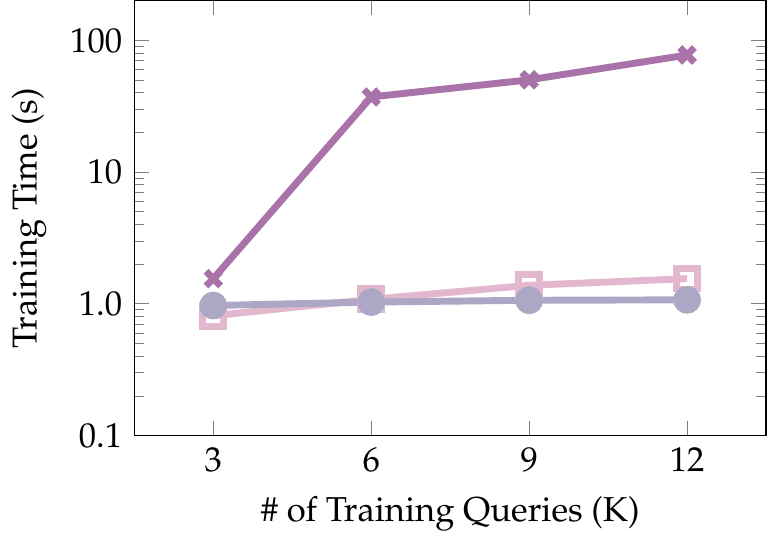}
}
%
\subfigure[\tpcds]{
\label{fig:exp:scale:memory:tpcds}
 \includegraphics[width=0.385\columnwidth]{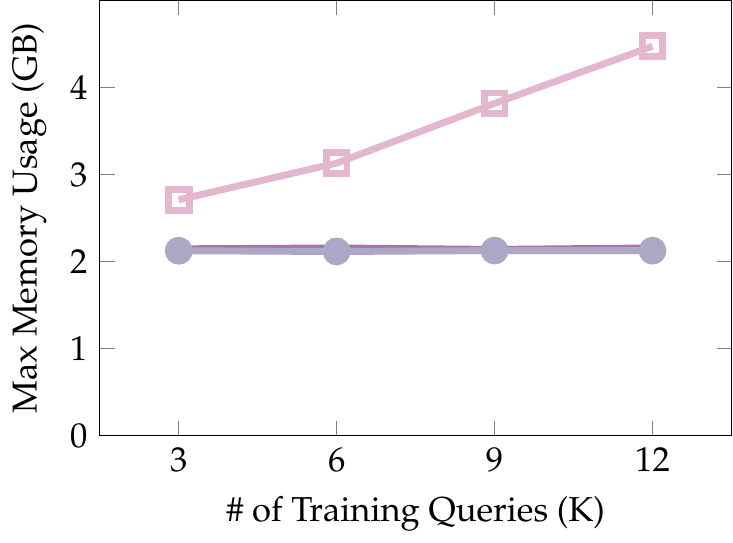}
}
\end{tabular}
\vspace*{-0.3cm}
\caption{Training Time \& Memory Usage}
\label{fig:exp:scale}
\end{figure}

\stitle{Scalability of Training}. We test the scalability of \NNGP estimator in the facets of training time and memory usage, comparing with the top-2 lightweight models \NN and \GBDT. 
As the the number of queries increases linearly,  Fig.~\ref{fig:exp:scale:time:forest},  \ref{fig:exp:scale:time:higgs},  \ref{fig:exp:scale:time:tpch} and \ref{fig:exp:scale:time:tpcds} show the training time on CPUs, where training an \NN takes tens to hundreds seconds while training \GBDT or \NNGP only takes less than 3 seconds.  
Although the training complexity of \NNGP is $O(N^3)$, the exact Bayesian inference is boosted by highly paralleled basic linear algebra operations. 
The time complexity of \NN is $O(TNd^{2}h)$, where $T$, $d$, and $h$ are the number of iterations/epochs, dimension of input and the number of neurons in the hidden layer.  
Extra overhead is involved in the forward/backward propagation. 
Fig.~\ref{fig:exp:scale:memory:forest}, \ref{fig:exp:scale:memory:higgs},  \ref{fig:exp:scale:memory:tpch} and  \ref{fig:exp:scale:memory:tpcds} show the peak memory usage monitored in the training phase.
Compared to \NN and \GBDT,  \NNGP needs more memory to persist the kernel matrix, which is quadratic to the number of training queries.  
Fortunately, as we will show in \cref{sec:exp:robust}, as a nonparametric model, \NNGP already achieves satisfying and robust performance under a small volume of training data.
It is worth noting that all the other estimators are more time and memory consuming than the two-layer \NN. 
Given the same epoch, training of \MSCN is roughly constant time slower than that of \NN and training of \TLSTM is even much slower than that of \MSCN and \NN. 
For the data-driven estimators, \DeepDBExp and \NeuroCardExp, the resources consumed are directly determined by the scale of the input relations.
Among all the learned estimators, the most consuming is \NeuroCardExp, where training on the large dataset \higgs is failed within 72 hours.

\subsection{Robustness to Workload Shifts}
\label{sec:exp:robust}
\begin{figure}[t]
\centering
\begin{tabular}[t]{c}
\includegraphics[width=0.3\columnwidth]{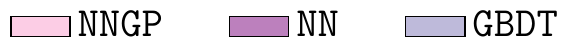}
\end{tabular}
\begin{tabular}[t]{c}
\subfigure[Varying the number of training queries for \forest]{
\label{fig:exp:robust:scale:forest}
 \includegraphics[width=0.9\columnwidth]{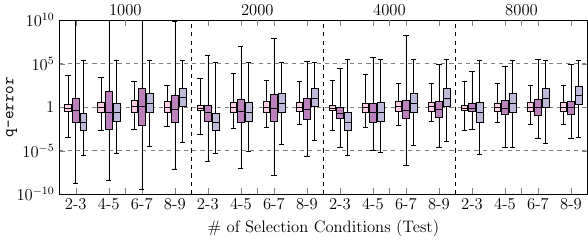}
}
\end{tabular}
\begin{tabular}[t]{c}
\subfigure[Varying the fraction of predicate size for \forest]{
\label{fig:exp:robust:skew:forest}
 \includegraphics[width=0.9\columnwidth]{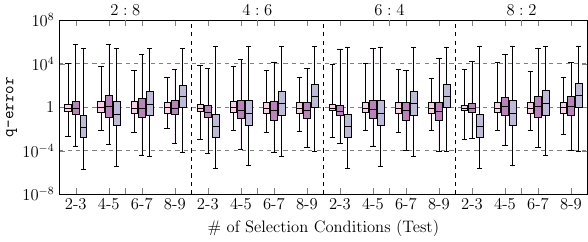}
}
\end{tabular}

\begin{tabular}[t]{c}
\subfigure[Varying the number of training queries for \tpch]{
\label{fig:exp:robust:scale}
 \includegraphics[width=0.9\columnwidth]{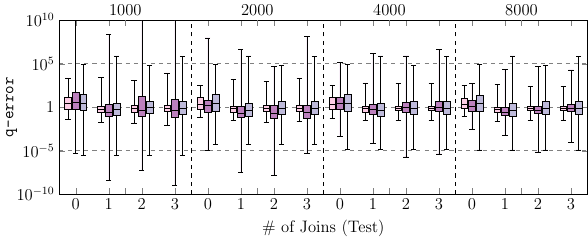}
}
\end{tabular}
\begin{tabular}[t]{c}
\subfigure[Varying the fraction of join size for \tpch]{
\label{fig:exp:robust:skew}
 \includegraphics[width=0.9\columnwidth]{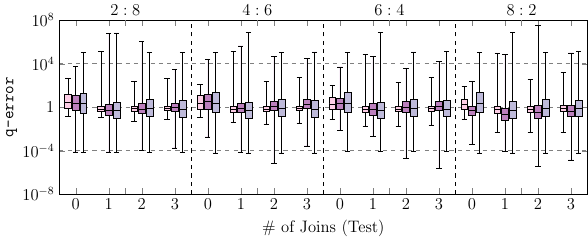}
}
\end{tabular}
\vspace*{-0.3cm}
\caption{Robustness to Various Training Workloads}
\label{fig:exp:robust}
\end{figure}

To study the robustness of \NNGP estimator, we train different models independently over various training workloads of \forest and \tpch, and test the models on split fixed test query sets, i.e., 20\% of the whole query sets.
The test queries are evenly distributed on the number of selection conditions for \forest or the number of join conditions for \tpch.
To generate various training workloads, first, we control the total number of training queries to $1,000$, $2,000$, $4,000$, $8,000$, respectively, where the queries with different numbers of selection/join conditions are evenly distributed, and the result is shown in Fig.~\ref{fig:exp:robust:scale:forest} and \ref{fig:exp:robust:scale}. 
Second, the total number of training queries is fixed to 40\% of the whole queries, and the fraction of queries with less/more selection or join conditions is set within \{2:8, 4:6, 6:4, 8:2\}.
\forest queries with 2-5 selection conditions are regarded as less conditions queries while others are more conditions.  
\tpch queries with 0-1 joins are regarded as less conditions queries while others are more conditions. 
The testing results on the fixed 20\% test queries are shown in Fig.~\ref{fig:exp:robust:skew:forest} and~\ref{fig:exp:robust:skew}. 

In general, the key observation is that the \NNGP estimator performs supremely robust on various workloads, regardless of the number of training queries and the fraction of different queries.  
In Fig.~\ref{fig:exp:robust:scale:forest} and \ref{fig:exp:robust:scale}, even though there are only $1,000$ training queries,  the \Qerror statistics of \NNGP varies within one order of magnitude compared with the model with $8,000$ training queries, and even better than \NN with $8,000$ training queries. 
In contrast, due to overfitting, the \Qerror of its counterpart \NN is degenerated drastically in the scenario of fewer training queries, which is up to $10^{10}$. 
As the fraction of queries with less/more selection or join conditions shifts in Fig.~\ref{fig:exp:robust:skew:forest} or  Fig.~\ref{fig:exp:robust:skew}, respectively, the performance of \NNGP also oscillates slightly. 
As a nonparametric model, \GBDT is relatively more robust than \NN but has a larger \Qerror. 
We notice that \GBDT underfits on \forest, whose prediction has a large bias on the test queries and even on the training queries. 
In the experiments, we also observe that the performance of \NN estimator is not stable during multiple runs, which is influenced by parameter initialization and stochastic optimization. 

\subsection{Uncertainty for Active Learning}
\label{sec:exp:active}

\begin{figure}[t]
\centering
\begin{tabular}[t]{c}
\includegraphics[width=0.4\columnwidth]{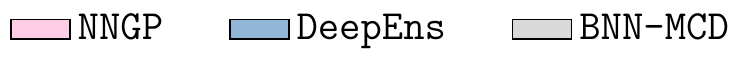}
\vspace*{-0.2cm} 
\end{tabular}
\begin{tabular}[t]{c}
\subfigure[\forest]{
\label{fig:exp:al:forest}
 \includegraphics[width=0.9\columnwidth]{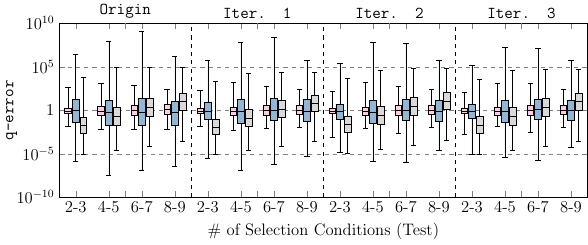}
}
\end{tabular}
\begin{tabular}[t]{c}
\subfigure[\tpch]{
\label{fig:exp:al:tpch}
 \includegraphics[width=0.9\columnwidth]{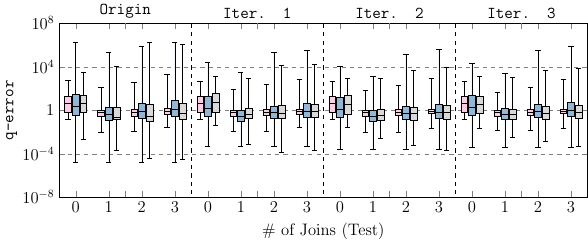}
}
\end{tabular}
\vspace*{-0.3cm}
\caption{\Qerror of 3 Iterations of Active Learning}
\label{fig:exp:al}
\end{figure}

\begin{table}[t]
\caption{mean-squared-error (MSE) on test queries}
\vspace*{-0.3cm}
\label{exp:tbl:active}
\centering
\begin{tabular}{|l l r r r r|}
\hline
{\bf MSE } & {\bf Estimator} & Origin & Iter. 1 & Iter. 2 & Iter. 3 \\ \hline
\hline
 \multirow{3}{*}{\forest} & \NNGP &  6.27  &  5.79 &  5.65 & 5.50 \\ \cline{2-6}
 						  & \DeepEns &  45.43  & 38.00 &  35.26 & 33.50 \\ \cline{2-6} 
						  & \BNNMCD &  36.18  &  33.63 &  34.87 & 34.80 \\
\hline
\hline
 \multirow{3}{*}{\tpch} & \NNGP &  5.30  & 5.13 & 5.02 & 4.95 \\ \cline{2-6}
 						  & \DeepEns &  14.14  & 11.29 &  10.87 & 10.49 \\ \cline{2-6} 
						  & \BNNMCD &  16.78  & 11.45 & 11.35 & 11.53 \\ 
\hline
\end{tabular}
\end{table}

Finally, we investigate leveraging the predictive uncertainty to improve the model explicitly by active learning. 
The key step of active learning is to select a set of informative test data to enrich the original training data and update the model~\cite {DBLP:series/synthesis/2012Settles}.
The predictive uncertainty, as the model's belief on its prediction, is an efficient and effective selection criterion. 
The corresponding active learning algorithm, a.k.a., uncertainty sampling~\cite{DBLP:conf/icml/LewisC94}, is to sample 
from the region of the data which has the most uncertainty regarding the current model, request the ground truth for these test data and 
retrain/update the model by original and added data. 
This process is repeated for several iterations under a specified stop condition, e.g., a given iteration number or sampling budget.
As the existing DL-based estimators do not deliver the predictive uncertainty, we used the two uncertainty-aware DL-based estimators introduced in \cref{sec:uncertainty}, \DeepEns and \BNNMCD, as the baselines. 

To conduct active learning, the entire queries are split into 40\% for training an original base estimator, 20\% for testing and 40\% as a selection pool. The base \DeepEns (5 \NN estimators for the ensemble) and \BNNMCD are trained by 50 epochs. We apply 3 iterations of uncertainty sampling where each iteration 1,000 queries are drawn without replacement regarding the coefficient of variation. 
\DeepEns and \BNNMCD are updated on their base model, and \NNGP is retrained from scratch. 
Table~\ref{exp:tbl:active} presents the mean-squared-error (MSE) (Eq.~(\ref{eq:loss:mse:square})) for the base model and those after the 3 iterations, on the test query set. \NNGP and the two DL baselines are able to improve themselves introspectively via uncertainty-based active learning.  
Among them \NNGP achieves the lowest test MSE. The MSE of \NNGP and \DeepEns are consistently reduced in the 3 iterations although the MSE of \NNGP is already small.
A close observation on the prediction accuracy of different \forest and \tpch queries is shown in Fig.~\ref{fig:exp:al}. 
Due to the neural network nature, the performance of \BNNMCD and \DeepEns improves significantly as more queries are incorporated into the training set.   
As the test MSE of origin \NNGP is small, its \Qerror reduction is not obvious compared with \BNNMCD and \DeepEns. 
But we still can observe an improvement on the queries with 2-3 joins in Fig.~\ref{fig:exp:al:tpch}.
The model updating has marginal effect for all the three estimators.

\section{Related Work}  
\label{sec:rw}
\stitle{Classical Cardinality Estimators.}
Cardinality/Selectively estimation of relational queries is studied over decades in the database area.
Early approaches propose multi-dimensional histograms to represent the joint probability distribution~\cite{DBLP:conf/vldb/PoosalaI97, DBLP:journals/vldb/GunopulosKTD05, DBLP:conf/sigmod/PoosalaIHS96}, where the assumption of attribute value independence is not necessary to be held. 
For join queries, join sampling algorithms~\cite{haas1999ripple, DBLP:conf/sigmod/0001WYZ16, DBLP:conf/sigmod/CaiBS19, DBLP:conf/sigmod/ZhaoC0HY18} are designed, which can be used to approximate cardinality and query results. 
These approaches face the risk of sampling failure in the intermediate join step, when the data distribution is complex. 
\cite{DBLP:conf/sigmod/GetoorTK01, DBLP:journals/pvldb/TzoumasDJ11} use Bayesian Network to estimate cardinality by modeling the joint probability distribution of all the attributes.
Structural learning is needed to explore a well-structured DAG to represent potential conditional independence of attribute.
Note Bayesian Network is different from Bayesian neural network. 
The former is a probabilistic graphical model based on the conditional independence assumption whereas the latter is the Bayesian DL model. 

\stitle{ML/DL for Cardinality Estimation and AQP.}
ML/DL models are exploited to perform cardinality estimation and AQP
for \rdbm.  We briefly review ML/DL approaches in
Table~\ref{tbl:exist_approaches}.
\DBEst~\cite{DBLP:conf/sigmod/MaT19} builds kernel density
estimation (KDE) models and tree-based regressors to conduct AQP.
\DeepDB~\cite{DBLP:journals/pvldb/HilprechtSKMKB20} adopts Sum-Product
Networks (SPNs) to learn the joint probability distribution of
attributes. An SQL query is complied to a product of expectations or
probability queries on the SPNs, where the product is based on
independence of the attributes.
\cite{DBLP:conf/icde/Thirumuruganathan20} uses deep generative model,
e.g., Variational Autoencoder (VAE) to model the joint probability
distribution of attributes, which only support analytical aggregate
query on a single table.  \cite{DBLP:journals/pvldb/KieferHBM17}
estimates multivariate probability distributions of a relation to
perform cardinality estimation by KDE.  Multiple joins can be
estimated by building the KDE estimator on pre-computed join result or
an estimation formula that leverages samples from the multiple
relations as well as the models.  Deep autoregressive model, e.g.,
Masked Autoencoder (MADE), is also adopted to learn the joint
probability distribution \cite{DBLP:journals/pvldb/YangLKWDCAHKS19,
  DBLP:conf/sigmod/HasanTAK020}, which decomposes the joint
distribution to a product of conditional distributions.  Kipf
et al. propose a multi-set convolutional neural network (MSCN) to
express query features using sets \cite{DBLP:conf/cidr/KipfKRLBK19}.
Anshuman et al.~\cite{DBLP:journals/pvldb/DuttWNKNC19} use
MLP and tree-based regressor to express
multiple attributes range queries.  Sun
et al.~\cite{DBLP:journals/pvldb/SunL19} extract the features of
physical query plan by Tree LSTM to estimate the query execution cost
as well as the cardinality.

\stitle{ML/DL for Databases.} In recent years, with the development of ML/DL techniques, ML/DL models are exploited to support multiple database applications. 
Various types of ML/DL models serve as a cost estimator of algorithms or query plans, which are to support applications of data partitioning~\cite{DBLP:conf/sigmod/FanJLLLXYYZ20}, index recommendation~\cite{DBLP:conf/sigmod/DingDM0CN19} and concurrency control~\cite{DBLP:journals/pvldb/ZhouSLF20}. 
\cite{DBLP:conf/sigmod/NathanDAK20, DBLP:conf/sigmod/KraskaBCDP18, DBLP:conf/sigmod/DingMYWDLZCGKLK20, DBLP:conf/sigmod/MarcusZK20} propose learned index structures, which learn a cumulative distribution function of the underlying data. 
\cite{DBLP:journals/pvldb/MarcusNMZAKPT19, DBLP:journals/corr/abs-1808-03196,  DBLP:conf/sigmod/MarcusZK20} design end-to-end learning-based query optimizers where \cite{DBLP:journals/pvldb/MarcusNMZAKPT19, DBLP:journals/corr/abs-1808-03196} optimize the binary join order and \cite{DBLP:conf/sigmod/MarcusZK20} generates the physical plans. Their approaches reformulate the dynamical programming problem of query optimization to Markov Decision Process and adopt different reinforcement learning (RL) algorithms to learn neural network models as the optimizer. 
\cite{DBLP:conf/sigmod/AkenPGZ17, DBLP:conf/sigmod/ZhangLZLXCXWCLR19, DBLP:conf/sigmod/KunjirB20, DBLP:journals/pvldb/DuanTB09} adopt ML/DL to tune the database configurations, where GP-based Bayesian optimization and RL algorithms are used respectively to conduct an online tuning in \cite{DBLP:conf/sigmod/AkenPGZ17, DBLP:journals/pvldb/DuanTB09} and \cite{DBLP:conf/sigmod/ZhangLZLXCXWCLR19, DBLP:conf/sigmod/KunjirB20}.
It is worth mentioning that this paper is the first exploration of Bayesian DL in the 
database area.

\section{Conclusion}
\label{sec:conclusion}

In this paper, we explore a new simple yet effective NNGP estimator to
estimate cardinality of SQL queries. We compare it with 7 baseline
estimators over 4 relational datasets. In terms of accuracy, NNGP is
one of the top-3 estimators, and performs best in many cases. In
terms of efficiency of training, NNGP is 1-2 orders faster than the baseline estimators except GBDT. 
GBDT is marginally more efficient than NNGP but has a low prediction
accuracy. In terms of uncertainty, NNGP can consistently improve its
accuracy by uncertainty sampling via active learning, whereas BNN cannot do so. And it 
achieves a much smaller prediction error comparing to \DeepEnsemble and BNN. In
addition, NNGP is supremely robust on various workloads, and can be
learned with much fewer training queries. Our source code is public available 
in \url{https://github.com/Kangfei/NNGP}.

\section*{Acknowledgement}
We thank Zongheng Yang, the author of
\NeuroCard~\cite{DBLP:journals/corr/abs-2006-08109} for his help in
testing the estimator.

{
\bibliographystyle{abbrv}
\pagestyle{plain} 
\bibliography{ref}

\begin{thebibliography}{10}

\bibitem{pytorch}
{Pytorch}.
\newblock \url{https://github.com/pytorch/pytorch}.

\bibitem{DBLP:conf/sigmod/AkenPGZ17}
D.~V. Aken, A.~Pavlo, G.~J. Gordon, and B.~Zhang.
\newblock Automatic database management system tuning through large-scale
  machine learning.
\newblock In {\em Proc. {SIGMOD}'17}, pages 1009--1024, 2017.

\bibitem{DBLP:books/lib/Bishop07}
C.~M. Bishop.
\newblock {\em Pattern recognition and machine learning, 5th Edition}.
\newblock Information science and statistics. Springer, 2007.

\bibitem{jax2018github}
J.~Bradbury, R.~Frostig, P.~Hawkins, M.~J. Johnson, C.~Leary, D.~Maclaurin,
  G.~Necula, A.~Paszke, J.~Vander{P}las, S.~Wanderman-{M}ilne, and Q.~Zhang.
\newblock {JAX}: composable transformations of {P}ython+{N}um{P}y programs,
  2018.

\bibitem{DBLP:conf/sigmod/CaiBS19}
W.~Cai, M.~Balazinska, and D.~Suciu.
\newblock Pessimistic cardinality estimation: Tighter upper bounds for
  intermediate join cardinalities.
\newblock In {\em Proc. {SIGMOD}'19}, pages 18--35, 2019.

\bibitem{DBLP:conf/nips/ChenSWJ17}
J.~Chen, M.~Stern, M.~J. Wainwright, and M.~I. Jordan.
\newblock Kernel feature selection via conditional covariance minimization.
\newblock In {\em Proc. {NIPS}'19}, pages 6946--6955, 2017.

\bibitem{caDBLP:conf/kdd/ChenG16}
T.~Chen and C.~Guestrin.
\newblock Xgboost: {A} scalable tree boosting system.
\newblock In {\em Proc. {SIGKDD}'16}, pages 785--794, 2016.

\bibitem{DBLP:conf/nips/ChoS09}
Y.~Cho and L.~K. Saul.
\newblock Kernel methods for deep learning.
\newblock In {\em Proc. {NIPS}'09}, pages 342--350, 2009.

\bibitem{DBLP:conf/sigmod/DingDM0CN19}
B.~Ding, S.~Das, R.~Marcus, W.~Wu, S.~Chaudhuri, and V.~R. Narasayya.
\newblock {AI} meets {AI:} leveraging query executions to improve index
  recommendations.
\newblock In {\em Proc. {SIGMOD}'19}, pages 1241--1258, 2019.

\bibitem{DBLP:conf/sigmod/DingMYWDLZCGKLK20}
J.~Ding, U.~F. Minhas, J.~Yu, C.~Wang, J.~Do, Y.~Li, H.~Zhang, B.~Chandramouli,
  J.~Gehrke, D.~Kossmann, D.~B. Lomet, and T.~Kraska.
\newblock {ALEX:} an updatable adaptive learned index.
\newblock In {\em Proc. {SIGMOD}'20}, pages 969--984, 2020.

\bibitem{Dua:2019}
D.~Dua and C.~Graff.
\newblock {UCI} machine learning repository, 2017.

\bibitem{DBLP:journals/pvldb/DuanTB09}
S.~Duan, V.~Thummala, and S.~Babu.
\newblock Tuning database configuration parameters with ituned.
\newblock {\em Proc. {VLDB} Endow.}, 2(1):1246--1257.

\bibitem{dutt2020efficiently}
A.~Dutt, C.~Wang, V.~Narasayya, and S.~Chaudhuri.
\newblock Efficiently approximating selectivity functions using low overhead
  regression models.
\newblock {\em Proc. VLDB Endow.}, 13(12):2215--2228, 2020.

\bibitem{DBLP:journals/pvldb/DuttWNKNC19}
A.~Dutt, C.~Wang, A.~Nazi, S.~Kandula, V.~R. Narasayya, and S.~Chaudhuri.
\newblock Selectivity estimation for range predicates using lightweight models.
\newblock {\em Proc. {VLDB}}, 12(9):1044--1057, 2019.

\bibitem{DBLP:conf/sigmod/FanJLLLXYYZ20}
W.~Fan, R.~Jin, M.~Liu, P.~Lu, X.~Luo, R.~Xu, Q.~Yin, W.~Yu, and J.~Zhou.
\newblock Application driven graph partitioning.
\newblock In {\em Proc. {SIGMOD}'20}, pages 1765--1779, 2020.

\bibitem{DBLP:journals/corr/GalG15a}
Y.~Gal and Z.~Ghahramani.
\newblock Bayesian convolutional neural networks with bernoulli approximate
  variational inference.
\newblock {\em CoRR}, abs/1506.02158, 2015.

\bibitem{DBLP:conf/icml/GalG16}
Y.~Gal and Z.~Ghahramani.
\newblock Dropout as a bayesian approximation: Representing model uncertainty
  in deep learning.
\newblock In {\em Proc. {ICML}'16}, volume~48, pages 1050--1059, 2016.

\bibitem{DBLP:conf/iclr/Garriga-AlonsoR19}
A.~Garriga{-}Alonso, C.~E. Rasmussen, and L.~Aitchison.
\newblock Deep convolutional networks as shallow gaussian processes.
\newblock In {\em Proc. {ICLR}'19}, 2019.

\bibitem{DBLP:conf/icml/GermainGML15}
M.~Germain, K.~Gregor, I.~Murray, and H.~Larochelle.
\newblock {MADE:} masked autoencoder for distribution estimation.
\newblock In {\em Proc. {ICML}'15}, volume~37, pages 881--889, 2015.

\bibitem{DBLP:conf/sigmod/GetoorTK01}
L.~Getoor, B.~Taskar, and D.~Koller.
\newblock Selectivity estimation using probabilistic models.
\newblock In {\em Proc. {SIGMOD}'01}, pages 461--472, 2001.

\bibitem{DBLP:journals/vldb/GunopulosKTD05}
D.~Gunopulos, G.~Kollios, V.~J. Tsotras, and C.~Domeniconi.
\newblock Selectivity estimators for multidimensional range queries over real
  attributes.
\newblock {\em {VLDB} J.}, 14(2):137--154, 2005.

\bibitem{DBLP:conf/icml/GuoPSW17}
C.~Guo, G.~Pleiss, Y.~Sun, and K.~Q. Weinberger.
\newblock On calibration of modern neural networks.
\newblock In {\em Proc. {ICML}'17}, volume~70, pages 1321--1330. {PMLR}, 2017.

\bibitem{haas1999ripple}
P.~J. Haas and J.~M. Hellerstein.
\newblock Ripple joins for online aggregation.
\newblock {\em ACM SIGMOD Record}, 28(2):287--298, 1999.

\bibitem{DBLP:conf/sigmod/HasanTAK020}
S.~Hasan, S.~Thirumuruganathan, J.~Augustine, N.~Koudas, and G.~Das.
\newblock Deep learning models for selectivity estimation of multi-attribute
  queries.
\newblock In {\em Proc. {SIGMOD}'20}, pages 1035--1050, 2020.

\bibitem{DBLP:journals/pvldb/HilprechtSKMKB20}
B.~Hilprecht, A.~Schmidt, M.~Kulessa, A.~Molina, K.~Kersting, and C.~Binnig.
\newblock Deepdb: Learn from data, not from queries!
\newblock {\em Proc. {VLDB}}, 13(7):992--1005, 2020.

\bibitem{DBLP:journals/nn/Hornik91}
K.~Hornik.
\newblock Approximation capabilities of multilayer feedforward networks.
\newblock {\em Neural Networks}, 4(2):251--257, 1991.

\bibitem{DBLP:journals/nn/HornikSW90}
K.~Hornik, M.~B. Stinchcombe, and H.~White.
\newblock Universal approximation of an unknown mapping and its derivatives
  using multilayer feedforward networks.
\newblock {\em Neural Networks}, 3(5):551--560, 1990.

\bibitem{DBLP:conf/icml/HronBSN20}
J.~Hron, Y.~Bahri, J.~Sohl{-}Dickstein, and R.~Novak.
\newblock Infinite attention: {NNGP} and {NTK} for deep attention networks.
\newblock In {\em Proc. {ICML}'20}, volume 119, pages 4376--4386. {PMLR}, 2020.

\bibitem{DBLP:journals/corr/abs-2002-12168}
J.~Hu, J.~Shen, B.~Yang, and L.~Shao.
\newblock Infinitely wide graph convolutional networks: Semi-supervised
  learning via gaussian processes.
\newblock {\em CoRR}, abs/2002.12168, 2020.

\bibitem{hutter2019automated}
F.~Hutter, L.~Kotthoff, and J.~Vanschoren.
\newblock {\em Automated machine learning: methods, systems, challenges}.
\newblock Springer Nature, 2019.

\bibitem{DBLP:journals/pvldb/KieferHBM17}
M.~Kiefer, M.~Heimel, S.~Bre{\ss}, and V.~Markl.
\newblock Estimating join selectivities using bandwidth-optimized kernel
  density models.
\newblock {\em Proc. {VLDB}}, 10(13):2085--2096, 2017.

\bibitem{DBLP:conf/cidr/KipfKRLBK19}
A.~Kipf, T.~Kipf, B.~Radke, V.~Leis, P.~A. Boncz, and A.~Kemper.
\newblock Learned cardinalities: Estimating correlated joins with deep
  learning.
\newblock In {\em Proc. {CIDR}'19}, 2019.

\bibitem{DBLP:conf/sigmod/KraskaBCDP18}
T.~Kraska, A.~Beutel, E.~H. Chi, J.~Dean, and N.~Polyzotis.
\newblock The case for learned index structures.
\newblock In {\em Proc. {SIGMOD}'18}, pages 489--504, 2018.

\bibitem{DBLP:journals/corr/abs-1808-03196}
S.~Krishnan, Z.~Yang, K.~Goldberg, J.~M. Hellerstein, and I.~Stoica.
\newblock Learning to optimize join queries with deep reinforcement learning.
\newblock {\em CoRR}, abs/1808.03196, 2018.

\bibitem{DBLP:conf/sigmod/KunjirB20}
M.~Kunjir and S.~Babu.
\newblock Black or white? how to develop an autotuner for memory-based
  analytics.
\newblock In {\em Proc. {SIGMOD}'20}, pages 1667--1683, 2020.

\bibitem{DBLP:conf/nips/Lakshminarayanan17}
B.~Lakshminarayanan, A.~Pritzel, and C.~Blundell.
\newblock Simple and scalable predictive uncertainty estimation using deep
  ensembles.
\newblock In {\em Proc. {NIPS}'17}, pages 6402--6413, 2017.

\bibitem{lee2004bayesian}
H.~K. Lee.
\newblock {\em Bayesian nonparametrics via neural networks}.
\newblock SIAM, 2004.

\bibitem{DBLP:conf/iclr/LeeBNSPS18}
J.~Lee, Y.~Bahri, R.~Novak, S.~S. Schoenholz, J.~Pennington, and
  J.~Sohl{-}Dickstein.
\newblock Deep neural networks as gaussian processes.
\newblock In {\em Proc. {ICLR}'18}, 2018.

\bibitem{DBLP:conf/icml/LewisC94}
D.~D. Lewis and J.~Catlett.
\newblock Heterogeneous uncertainty sampling for supervised learning.
\newblock In {\em Proc. {ICML}'94}, pages 148--156, 1994.

\bibitem{DBLP:conf/sigmod/0001WYZ16}
F.~Li, B.~Wu, K.~Yi, and Z.~Zhao.
\newblock Wander join: Online aggregation via random walks.
\newblock In {\em Proc. {SIGMOD}'16}, pages 615--629, 2016.

\bibitem{DBLP:journals/corr/abs-1903-01363}
X.~Liang, A.~J. Elmore, and S.~Krishnan.
\newblock Opportunistic view materialization with deep reinforcement learning.
\newblock {\em CoRR}, abs/1903.01363, 2019.

\bibitem{DBLP:conf/sigmod/MaT19}
Q.~Ma and P.~Triantafillou.
\newblock Dbest: Revisiting approximate query processing engines with machine
  learning models.
\newblock In {\em Proc. {SIGMOD}'19}, pages 1553--1570, 2019.

\bibitem{mackay1998introduction}
D.~J. MacKay.
\newblock Introduction to gaussian processes.
\newblock {\em NATO ASI series F computer and systems sciences}, 168:133--166,
  1998.

\bibitem{DBLP:conf/sigmod/MarcusZK20}
R.~Marcus, E.~Zhang, and T.~Kraska.
\newblock Cdfshop: Exploring and optimizing learned index structures.
\newblock In {\em Proc. {SIGMOD}'20}, pages 2789--2792, 2020.

\bibitem{DBLP:journals/pvldb/MarcusNMZAKPT19}
R.~C. Marcus, P.~Negi, H.~Mao, C.~Zhang, M.~Alizadeh, T.~Kraska,
  O.~Papaemmanouil, and N.~Tatbul.
\newblock Neo: {A} learned query optimizer.
\newblock {\em Proc. {VLDB} Endow.}, 12(11):1705--1718, 2019.

\bibitem{DBLP:conf/aaai/NathD15}
A.~Nath and P.~M. Domingos.
\newblock Learning relational sum-product networks.
\newblock In {\em Proc. {AAAI}'15}, pages 2878--2886, 2015.

\bibitem{DBLP:conf/sigmod/NathanDAK20}
V.~Nathan, J.~Ding, M.~Alizadeh, and T.~Kraska.
\newblock Learning multi-dimensional indexes.
\newblock In {\em Proc. {SIGMOD}'20}, pages 985--1000, 2020.

\bibitem{neal1996priors}
R.~M. Neal.
\newblock Priors for infinite networks.
\newblock In {\em Bayesian Learning for Neural Networks}, pages 29--53.
  Springer, 1996.

\bibitem{DBLP:conf/iclr/NovakXBLYHAPS19}
R.~Novak, L.~Xiao, Y.~Bahri, J.~Lee, G.~Yang, J.~Hron, D.~A. Abolafia,
  J.~Pennington, and J.~Sohl{-}Dickstein.
\newblock Bayesian deep convolutional networks with many channels are gaussian
  processes.
\newblock In {\em Proc. {ICLR}'19}, 2019.

\bibitem{neuraltangents2020}
R.~Novak, L.~Xiao, J.~Hron, J.~Lee, A.~A. Alemi, J.~Sohl-Dickstein, and S.~S.
  Schoenholz.
\newblock Neural tangents: Fast and easy infinite neural networks in python.
\newblock In {\em Proc. {ICLR}'20}, 2020.

\bibitem{DBLP:conf/vldb/PoosalaI97}
V.~Poosala and Y.~E. Ioannidis.
\newblock Selectivity estimation without the attribute value independence
  assumption.
\newblock In {\em Proc. {VLDB}'97}, pages 486--495, 1997.

\bibitem{DBLP:conf/sigmod/PoosalaIHS96}
V.~Poosala, Y.~E. Ioannidis, P.~J. Haas, and E.~J. Shekita.
\newblock Improved histograms for selectivity estimation of range predicates.
\newblock In {\em Proc. {SIGMOD}}, pages 294--305, 1996.

\bibitem{DBLP:books/lib/RasmussenW06}
C.~E. Rasmussen and C.~K.~I. Williams.
\newblock {\em Gaussian processes for machine learning}.
\newblock {MIT} Press, 2006.

\bibitem{DBLP:series/synthesis/2012Settles}
B.~Settles.
\newblock {\em Active Learning}.
\newblock Synthesis Lectures on Artificial Intelligence and Machine Learning.
  Morgan {\&} Claypool Publishers, 2012.

\bibitem{DBLP:journals/jmlr/SongSGBB12}
L.~Song, A.~J. Smola, A.~Gretton, J.~Bedo, and K.~M. Borgwardt.
\newblock Feature selection via dependence maximization.
\newblock {\em J. Mach. Learn. Res.}, 13:1393--1434, 2012.

\bibitem{DBLP:journals/pvldb/SunL19}
J.~Sun and G.~Li.
\newblock An end-to-end learning-based cost estimator.
\newblock {\em Proc. {VLDB}}, 13(3):307--319, 2019.

\bibitem{DBLP:conf/acl/TaiSM15}
K.~S. Tai, R.~Socher, and C.~D. Manning.
\newblock Improved semantic representations from tree-structured long
  short-term memory networks.
\newblock In {\em Proc. {ACL}'15}, pages 1556--1566, 2015.

\bibitem{DBLP:conf/icde/Thirumuruganathan20}
S.~Thirumuruganathan, S.~Hasan, N.~Koudas, and G.~Das.
\newblock Approximate query processing for data exploration using deep
  generative models.
\newblock In {\em Proc. {ICDE}'20}, pages 1309--1320, 2020.

\bibitem{DBLP:journals/pvldb/TzoumasDJ11}
K.~Tzoumas, A.~Deshpande, and C.~S. Jensen.
\newblock Lightweight graphical models for selectivity estimation without
  independence assumptions.
\newblock {\em Proc. {VLDB} Endow.}, 4(11):852--863, 2011.

\bibitem{DBLP:conf/nips/VaswaniSPUJGKP17}
A.~Vaswani, N.~Shazeer, N.~Parmar, J.~Uszkoreit, L.~Jones, A.~N. Gomez,
  L.~Kaiser, and I.~Polosukhin.
\newblock Attention is all you need.
\newblock In {\em Proc. {NeurIPS}'17}, pages 5998--6008, 2017.

\bibitem{DBLP:journals/tkde/WangY16}
H.~Wang and D.~Yeung.
\newblock Towards bayesian deep learning: {A} framework and some existing
  methods.
\newblock {\em {IEEE} Trans. Knowl. Data Eng.}, 28(12):3395--3408, 2016.

\bibitem{DBLP:journals/corr/abs-2012-06743}
X.~Wang, C.~Qu, W.~Wu, J.~Wang, and Q.~Zhou.
\newblock Are we ready for learned cardinality estimation?
\newblock {\em CoRR}, abs/2012.06743, 2020.

\bibitem{DBLP:journals/neco/Williams98}
C.~K.~I. Williams.
\newblock Computation with infinite neural networks.
\newblock {\em Neural Comput.}, 10(5):1203--1216, 1998.

\bibitem{DBLP:conf/nips/WilsonI20}
A.~G. Wilson and P.~Izmailov.
\newblock Bayesian deep learning and a probabilistic perspective of
  generalization.
\newblock In {\em Proc. {NeurIPS}'20}, 2020.

\bibitem{DBLP:conf/nips/Yang19}
G.~Yang.
\newblock Wide feedforward or recurrent neural networks of any architecture are
  gaussian processes.
\newblock In {\em Proc. {NeurIPS}'19}, pages 9947--9960, 2019.

\bibitem{DBLP:journals/corr/abs-2006-08109}
Z.~Yang, A.~Kamsetty, S.~Luan, E.~Liang, Y.~Duan, P.~Chen, and I.~Stoica.
\newblock Neurocard: One cardinality estimator for all tables.
\newblock {\em Proc. {VLDB} Endow.}, 14(1):61--73, 2020.

\bibitem{DBLP:journals/pvldb/YangLKWDCAHKS19}
Z.~Yang, E.~Liang, A.~Kamsetty, C.~Wu, Y.~Duan, P.~Chen, P.~Abbeel, J.~M.
  Hellerstein, S.~Krishnan, and I.~Stoica.
\newblock Deep unsupervised cardinality estimation.
\newblock {\em Proc. {VLDB}}, 13(3):279--292, 2019.

\bibitem{DBLP:conf/sigmod/ZhangLZLXCXWCLR19}
J.~Zhang, Y.~Liu, K.~Zhou, G.~Li, Z.~Xiao, B.~Cheng, J.~Xing, Y.~Wang,
  T.~Cheng, L.~Liu, M.~Ran, and Z.~Li.
\newblock An end-to-end automatic cloud database tuning system using deep
  reinforcement learning.
\newblock In {\em Proc. {SIGMOD}'19}, pages 415--432, 2019.

\bibitem{DBLP:conf/sigmod/ZhaoC0HY18}
Z.~Zhao, R.~Christensen, F.~Li, X.~Hu, and K.~Yi.
\newblock Random sampling over joins revisited.
\newblock In {\em Proc. {SIGMOD}'18}, pages 1525--1539, 2018.

\bibitem{DBLP:journals/pvldb/ZhouSLF20}
X.~Zhou, J.~Sun, G.~Li, and J.~Feng.
\newblock Query performance prediction for concurrent queries using graph
  embedding.
\newblock {\em Proc. {VLDB} Endow.}, 13(9):1416--1428, 2020.

\end{thebibliography}
}


\end{document}